\newcommand{\Label}[1]{\label{#1}}
\newcommand{\bal}[1]{\begin{eqnarray}\label{#1}}
\newcommand{\bel}[1]{\begin{equation}\label{#1}}

\newcommand{\eqn}[1]{(\ref{#1})}
\newcommand{\eqnn}[2]{(\ref{#1}, \ref{#2})}
\newcommand{\eqnnn}[3]{(\ref{#1}, \ref{#2}, \ref{#3})}

\newcommand{\be}{\begin{equation}}
\newcommand{\ee}{\end{equation}}
\newcommand{\ba}{\begin{eqnarray}}
\newcommand{\ea}{\end{eqnarray}}
\newcommand{\nn}{\nonumber\\}

\newcommand{\intinf}{\int_{-\infty}^\infty}
\newcommand{\inta}{\int_{z_i}^\lambda}
\newcommand{\intb}{\int_\lambda^z}

\newcommand{\braket}[1]{\langle #1 \rangle}

\newcommand{\ls}{<_{_{_{\!\!\!\!\!\!\!\textstyle{\sim}}}}}

\newcommand{\re}{\mbox{Re }}
\newcommand{\im}{\mbox{Im }}

\newcommand{\sh}{\;\mbox{sh}\,}
\newcommand{\ch}{\;\mbox{ch}\,}
\newcommand{\th}{\;\mbox{th}\,}
\newcommand{\shsq}{\;\mbox{sh}^2\,}

\newcommand{\tr}{\mbox{tr }}

\newcommand{\bold}[1]{\mbox{\boldmath $#1$}} 

\newskip\humongous \humongous=0pt plus 1000pt minus 1000pt

\newif\ifdtup

\documentstyle[12pt,epsfig]{article}

\makeatletter                    
\@addtoreset{equation}{section}  
\makeatother                     

\begin{document}
\title{Entropy and Uncertainty of Squeezed Quantum Open Systems}

\author{Don Koks \thanks{e-mail address: dkoks@physics.adelaide.edu.au}\\
{\small Department of Physics and Mathematical Physics,}\\
{\small University of Adelaide,
Adelaide SA 5005, Australia}\\
Andrew Matacz \thanks{e-mail address: andrewm@maths.su.oz.au}\\
{\small School of Mathematics and Statistics, University of Sydney, 
Sydney NSW 2006, Australia}\\
and B. L. Hu \thanks{e-mail address: hu@umdhep.umd.edu}\\
{\small Department of Physics, University of Maryland,
College Park, MD 20742, USA }}
\date{{\small {\it ADP-96-42/M50, umdpp 97-63, School of Mathematics and 
Statistics 96-41; Submitted to Physical Review D on Nov. 29, 1996
}}}
\maketitle

\begin{abstract}
We define  the  entropy $S$ and uncertainty function of a squeezed system
interacting with a thermal bath, and study how they change in time by
following the evolution of the reduced density matrix in the influence
functional formalism. As examples, we calculate the entropy of two
exactly solvable squeezed systems: an inverted harmonic oscillator
and a scalar field mode evolving in an inflationary universe.
For the inverted oscillator with weak coupling to the bath,
at both high and low temperatures, $S\to r $, where
$r$ is the squeeze parameter. In the de Sitter case, at high temperatures,
$S\to (1-c)r$ where $c = \gamma_0/H$, $\gamma_0$ being the coupling to the bath
and $H$  the Hubble constant.
These three cases confirm previous results
based on more ad hoc prescriptions for calculating entropy. But at
low temperatures, the de Sitter entropy  $S\to (1/2-c)r$ is noticeably 
different. This result, obtained from a more rigorous approach, 
shows that factors usually ignored by the
conventional approaches, i.e., the nature of the environment and the coupling
strength betwen the system and the environment, are important.

\end{abstract}

\newpage
\section{Introduction}

In discussing the conceptual problems of entropy generation from cosmological
particle creation~\cite{Par69,Zel70} one of us was confronted in the early 
80's~\cite{Hu82} by the following apparent paradox: on the one hand common 
sense
suggests that entropy ($S$) is given by the number ($N$) of particles produced
($S \approx N^3$ for photons). On the other hand, theoretically, for a free
field, particle pairs
created in the vacuum will remain in a pure state and there should be no
entropy generation. Inquiry into this paradox led to serious subsequent
investigations into the statistical properties of particles and fields.
In 1984, Hu~\cite{Hu84} pointed out that the usual simplistic identification
of entropy with the number of  particles present is valid
only in the thermodynamic-hydrodynamic regime, where interaction among
particles and coarse-graining can lead to entropy generation.
This aspect was discussed
later by Hu and Kandrup~\cite{HuKan} using a statistical mechanics
subdynamics analysis. The more intriguing case of entropy generation for
free fields was addressed by Hu and Pavon~\cite{HuPav}. They suggested
that an intrinsic entropy of a (free) quantum field can be measured by the
particle number (in a Fock space representation) or by the variance (in the
coherent state representation). The entropy of a (free) quantum field
is non-zero only if some information of the field is lost or excluded from
consideration, either by choosing some  special initial state and/or
introducing some measure of coarse-graining. For example, the predicted
monotonic increase in the spontaneous creation of bosons is a consequence of
adopting the Fock space representation which amounts to a random phase
initial condition implicitly assumed in most discussions of
vacuum particle creation.
(The difference of spontaneous and stimulated creation of bosons versus 
fermions
was first pointed out by Parker~\cite{Par69}, and discussed in squeezed state
language by Hu, Kang and Matacz~\cite{HKM}). The relation of
random phase and particle creation was further elaborated by Kandrup
\cite{Kan88}.

Following these early discussions of the theoretical meaning of entropy of
quantum fields, a recent surge of interest on this issue  was stimulated
by the work of Brandenberger, Mukhanov and Prokopec (BMP)~\cite{BMP},
Gasperini and Giovannini (GG)~\cite{GasGio} and others
on the entropy content of primordial gravitons. The language of squeezed states
for the description of cosmological
particle creation was introduced by Grishchuk and Sidorov \cite{GriSid}.
Though the physics is the same~\cite{HKM,AFJP} as originally described by
Parker~\cite{Par69} and Zeldovich~\cite{Zel70}, the language brings closer
the comparison with similar problems in quantum optics, which shares
many interesting theoretical and practical issues \cite{Knight}. BMP
suggested a coarse-graining  of the field by integrating out the rotation
angles in the probability functional, while GG
considered a squeezed vacuum in terms of new variables which give the
maximum and minimum fluctuations, and suggested a coarse-graining by neglecting
information about the subfluctuant variable. Keski-Vakkuri studied entropy
generation from particle creation with many particle mixed initial 
states~\cite{Kes}.
Matacz~\cite{Mat} considered a squeezed vacuum of a harmonic oscillator system
with time-dependent frequency, and, motivated by the special role of coherent
states, modeled the effect of the environment by decohering the squeezed
vacuum in the coherent state representation.
Kruczenski, Oxman and Zaldarriaga~\cite{KOZ} also used a procedure of
setting off-diagonal elements in the density matrix  to zero before calculating
the entropy. Despite the variety of coarse-graining measures used, in the large
squeezing limit (late times) these approaches all give an entropy of
$S= 2r$ per mode, where $r$ is the squeezing parameter. This result which
gives the number of particles created at late times agrees with that obtained
in the original work of Hu and Pavon~\cite{HuPav}.

Noteworthy in this group of work is that the representation of the state of
the quantum field and the coarse-graining in the field are stipulated,
not derived. What is implicitly assumed or grossed over in these approaches
is the important process of decoherence -- the diminution
of the off-diagonal components of a reduced density matrix in a certain basis.
It is  a necessary condition for realizing the quantum to classical
transition \cite{envdec}. The deeper issues are to show explicitly
how entropy of particle creation depends
on the choice of specific initial state and/or  particular ways of
coarse-graining, and to understand how natural or plausible these choices of 
the
initial state representation or the coarse-graining measure are
in different realistic physical conditions \cite{HuSpain}.%
\footnote{This includes conditions when, for example, the quantum field is at
a finite temperature or is in disequilibrium, interacting with
other fields, or that its vacuum state is dictated by some natural choice,
e.g., in the earlier quantum cosmology regime such as
the Hartle-Hawking boundary condition leading to the Bunch-Davies vacuum
in de Sitter spacetime.}
To answer these questions, one needs to work with a more basic theoretical
framework, that of statistical mechanics of quantum fields. In recent years
we have approached the decoherence and entropy /uncertainty issues with
the quantum open system concept~\cite{qos} and the influence functional 
formalism~\cite{if}.
The purpose of this paper is to study the entropy and uncertainty
of quantum fields using the statistical mechanics of squeezed quantum open
systems as illustrated by quantum Brownian motion models.

In the quantum Brownian motion paradigmic depiction of quantum field theory
studied in the series of papers by Hu, Paz, Zhang~\cite{HPZ1,HPZ2}
and Hu and Matacz~\cite{HM2}, the system represented by the Brownian particle
can act as a detector
(as in the influence functional derivation of Unruh and Hawking radiation
\cite{Ang,RHA}),
a particular mode of a quantum  field (such as the homogeneous inflaton
field), or the scale factor of the background spacetime (as in
minisuperspace quantum cosmology), while the bath could be a set of coupled
oscillators, a quantum field, or just the high frequency sector of the field,
as in stochastic inflation. The statistical properties of the system are
depicted by the reduced density matrix (rdm) formed by integrating out the
details of the bath. One can use the rdm or the associated Wigner function
to calculate the statistical average of physical observables of the system,
such as  the uncertainty or the entropy functions.
The von Neumann entropy of an open system is then
\bel{von-neumann-entropy} 
S \equiv -\tr \rho_{red}\,\ln \rho_{red}
\ee
The uncertainty function measures
the effects of  vacuum and thermal fluctuations in the environment
(at zero and finite temperature) on the observables of the system
\cite{HuZhaUncer,AndHal}. The increase
of their variances due to these fluctuations gives rise to the uncertainty and
entropy increase.       The time-dependence of the
uncertainty function of an open system measures the varying relative
importance of thermal and vacuum fluctuations and their roles
in bringing about the decoherence of the system and the emergence of classical
behavior~\cite{HuZhaUncer,AndHal}.

The entropy function constructed from the reduced density matrix
(or the Wigner function) of a particular state measures the information loss
of the system in that state to the environment (or, in the
phraseology of~\cite{ZHP}, the `stability' characterized
by the loss of predictive power relative to the classical description).
One can study the entropy increase for a specific
state, or compare the entropy at each time for a variety of states characterized
by the squeeze parameter. The time scale of entropy increase, when entropy
arises from particle creation from the vacuum, should be comparable to
the decoherence time, which, for a high temperature bath, is very short.
Interaction with the environment also changes its dynamics from strictly unitary
to dissipative, the energy loss being measured by the viscosity function, which
governs the relaxation of the system into equilibrium with the environment.
The entropy function for such open systems
can also be used~\cite{AndHal,ZHP} as a measure of how close
different quantum states can lead to a classical dynamics.
For example, the coherent state being the state of
minimal uncertainty has the smallest entropy function~\cite{ZHP}
and a squeezed state in general has a greater uncertainty function
\cite{HuZhaUncer}. One can thus use the uncertainty to measure how classical
or `nonclassical' a quantum state is.

Using this first-principle approach for the calculation of the entropy function
leads to  more reliable results. With regard to the issue of entropy of
quantum fields raised at the beginning, we can now ask, what is the difference
of our more vigorous definition and that defined earlier with more ad hoc
prescriptions?

Foremost, the differences in design are obvious: the entropy of
~\cite{HuPav,BMP,GasGio}
and others refers to that of the field, and is obtained by coarse-graining
some information of the field itself, such as making a random phase
approximation, adopting the number basis, or integrating over the rotation
angles. The entropy of~\cite{HuZhaUncer,AndHal,ZHP}
refers to that of the open system and is obtained by coarse-graining
the environment. Why is it that for certain generic models in some common
limit (late time, high squeezing), both groups of work obtain the same result?
Under what conditions would they differ? Understanding this relation could
provide a more solid theoretical foundation for the intuitively-argued
definitions of field entropy.

At the formal level, supposing we have some system which has been decomposed
into two subsystems,  it can be shown~\cite{Page} that between
the entropies $S_1, S_2$ of the two subsystems, and that of the total system,
$S_{12}$, a triangle inequality holds:
\be
|S_1-S_2| \le S_{12} \le S_1+S_2
\ee
In particular, if the total system is closed and so in a pure state, then it 
has zero entropy, so that the two subsystems necessarily have equal 
entropies.\footnote{
This could be the reason why the derivation of black hole entropy
(see the recent review of Bekenstein~\cite{BekMG7}) can be
obtained equivalently by computing the entropy of the radiation
(e.g.,~\cite{FroNov}) emitted by the black hole, or by counting the
internal states (if one knows how!) of the black hole (e.g.,~\cite{bhstates}).
Physically one
can view what happens to the particle as a probe into the state of the field.
The application of open-system concepts to black hole entropy is a very 
fruitful avenue \cite{bhentropy}.}
Hence, asking for the entropy change of a system is equivalent to
asking for the entropy change of the environment it couples to, if the
overall closed system is in a pure state. Now consider the case of the
system as a detector (or a single mode of a field) and the environment as the
field. The information
lost in coarse-graining the field which was used to define the field entropy
in the above examples is precisely the information lost as registered
in the particle detector, which shows up in the calculaton of entropy from
the reduced density matrix. The bilinear coupling between the system
and the bath as used in the simple quantum Brownian motion models also ensures
that the
information registered in both sectors are directly commutable. This explains
the commonalities.  However, not all coarse-graining and coupling will lead
to the same results, as we shall explicitly demonstrate in some examples.

Another important feature of the entropy function obtained in our present
investigation which is not at all
clear in earlier studies is that it depends nonlocally on the entire
history of the squeezing parameter. This can be seen from the fact that
the rate of particle creation varies in time and its effect is history
dependent~\cite{HH80,CH87}. Existing methods of calculating the entropy
generation give results which only depend on the squeezing parameter at the
time when a particular coarse-graining (or dropping the off-diagonal components
of the density matrix) is implemented. These {\it ad hoc} choices
(of coarse-graining and the time it is introduced)
affect the generality of the earlier results.



The plan of this paper is as follows.  In Sec.~2 we give a brief summary of
a squeezed quantum system, using a general oscillator Hamiltonian as an
example. The notation is that of~\cite{HKM,HM2}. In Sec.~3 we give a brief
summary of open quantum systems in terms of influence functionals
\cite{if}, following the treatment of~\cite{HPZ1,HM2}. Readers familiar with
these background material can go directly to Sec.~4, which contains the
central material for the derivation of entropy and uncertainty functions.
In Sec.~5 we apply these
formulas to an oscillator system, recovering en route the earlier results of
\cite{HuZhaUncer,AndHal} for uncertainty at finite temperature, and of
\cite{ZHP} on entropy of coherent states. In Sec.~6 we apply our result to the
consideration of a scalar field in a de Sitter universe. We show the conditions
where one recovers the $S=2r$ result of all previous work, and more significantly,
the cases when they differ. We give a short discussion of our findings in
Sec.~7. The Appendices contain details of derivations.

%

\section{Squeezed Systems}

\subsection{Squeezed states and  density matrices}


Consider the general oscillator Hamiltonian
\bel{qbm3-b.1}
H(t) = f(t){a^2\over 2} + f^\ast(t){{a^\dagger}^2\over 2}+h(t)(a^\dagger a+1/2)
+d(t) a + d^\ast(t) a^\dagger + g(t)
\ee
where $d, f, g, h$ are arbitrary functions of time.
The propagator for this has been calculated in~\cite{HM2} and is
\bel{qbm3-b.15}
U(t,t_i) = S(r,\phi)R(\theta)D(p)e^{w-|p|^2/2}
\ee
where $p,w$ are defined in terms of the coefficients appearing in $H$, and
\bal{qbm3-b.12-and-b.16}
D(p) &=& \exp (-p^\ast a-{\rm h.c.}) \nn
R(\theta) &=& \exp -i\theta(a^\dagger a+1/2) \nn
S(r,\phi) &=& \exp (re^{-2i\phi}a^2/2-{\rm h.c.})
\ea
are the displacement, rotation and squeeze operators~\cite{HKM} respectively.
Suppose we start with a simple harmonic oscillator with lagrangian
\be
L={M\over 2}\left(\dot x^2-\Omega^2x^2\right)
\ee
If we construct a gaussian state in the position basis, with initially the same
width~$\sigma_0$ as that of the ground state of such an oscillator, displaced 
by some 
arbitrary amount and with a phase proportional to $x$, we find this to be
an eigenstate of the lowering operator, and is called a coherent state.  
Suppose we locate the
point $(\braket{x},\braket{p})$ in phase space and draw an ellipse about
this point, the lengths of whose axes being the uncertainties $\Delta x^2, 
\Delta p^2$.  Then as the oscillator evolves this uncertainty ellipse
revolves about the origin with angular speed~$\Omega$.  

A squeezed state is again such a state, but with an arbitrary initial 
width~$\sigma$.  We find that as the oscillator evolves the uncertainty ellipse
again revolves about the origin, but its axes change length and it can also
rotate about its own centre.

It turns out that the squeeze parameter $r$ is related to the width of such
a state:
\bel{nieto-7.6} 
r = \ln {\sigma_0\over \sigma}\;\;\;,\;\;\;
\sigma_0 \equiv \sqrt{\hbar\over2M\Omega}
\ee
Hence a coherent state has $r=0$, or zero squeezing.  A gaussian that initially
has a width smaller than $\sigma_0$ will evolve to a squeezed state with
some $r>0$.  We can generate a
squeezed state by applying $S(r,\phi)$ to the ground state of the simple
oscillator.  Consider the new operator
\be
b = U^\dagger aU \equiv \alpha \,a+\beta^\ast \,a^\dagger
\ee
where it turns out that
\bal{qbm3-b.10}
\alpha &=& e^{-i\theta} \ch r \nn
\beta  &=& -e^{-i(\theta+2\phi)} \sh r
\ea
Going from $a$ to $b$ is then just a Bogoliubov transformation, and so $\alpha,
\beta$ become Bogoliubov coefficients for our system.
Their equations of motion are
\bal{qbm3-b.13}
\dot \alpha &=& -ih\alpha -i f^\ast \beta \nn
\dot \beta &=& if\alpha + ih\beta
\ea
$$
\alpha(t_i) =1 \;\;\;,\;\;\; \beta(t_i) =0
$$
where $f,h$ as defined in the hamiltonian \eqn{qbm3-b.1} are calculated from
the general system lagrangian.  This lagrangian has time dependent mass and 
frequency, and we will also allow it to have a time dependent cross term 
denoted $2{\cal E}(t)$:
\bel{4-3}
L = {M(t)\over 2}\left(\dot x^2 + 2{\cal E}(t)\dot xx - \Omega^2(t) x^2\right)
\ee
Then $f,h$ are given by~\cite{HM2}
\bal{ent-4-4}
f &=& {1\over 2}\left[{M\over \kappa}(\Omega^2+{\cal E}^2)-{\kappa\over M}+
2i{\cal E}\right] \nn
h &=& {1\over 2}\left[{M\over \kappa}(\Omega^2+{\cal E}^2)+{\kappa\over M}
\right]
\ea
and $\kappa$ is an arbitrary positive constant that can be chosen to simplify
the relevant equations. 

In the next section we shall find that the quantity of much importance to our
work turns out to be the sum of the Bogoliubov coefficients, 
$X\equiv \alpha+\beta$.
It follows from~\eqn{qbm3-b.13} that $X$ satisfies the classical equation of 
motion for the system:
\bel{5-1}
\ddot X + {\dot M\over M}\dot X +\left(\Omega^2+{\dot {\cal E}}+{\dot M 
{\cal E}\over M}\right)X = 0
\ee
with initial conditions
\bel{3.751}
X(t_i) = 1 \;\;\; ; \;\;\; \dot X(t_i) = {-i\kappa\over M(t_i)} -{\cal E}(t_i)
\ee
With this result,  the usual task of finding the
Bogoliubov coefficients~$\alpha,\beta$ from two coupled first order 
differential equations is  reduced to that of solving one second order
equation for~$X$.

\subsection{Squeezing an inverted harmonic oscillator}

For an inverted oscillator, i.e.\ one with $\Omega^2 < 0$, at late times $r$ is
expected to blow up.  In that case we can calculate it from~\eqn{qbm3-b.10} as 
follows.                                                           h
\bel{3.870.8-4}
|\alpha| \to |\beta| \to e^r/2
\ee
so that
\bel{3.870.8-4.5}
r \to \ln(2|\alpha|)
\ee
Rather than use \eqn{qbm3-b.13} to calculate $\alpha$, once
we have $X$ we can extract $\alpha$ from it.  This is done by writing,
from~\eqn{qbm3-b.13},
\bal{3.752.3-1-and-2}
X &=& \alpha + \beta \nn
\dot X &=& i(f-h)\alpha + i(h-f^\ast)\beta
\ea
and solving for $\alpha, \beta\,$ using \eqn{ent-4-4}:
\bel{3.752.3.1-1}
\left\{{}^\alpha_\beta\right\} = {1\over 2}\left(1\pm {i{\cal E}M\over \kappa}
\right)X \pm {iM\over 2\kappa}\dot X
\ee

We can follow the behaviour of $r, \phi, \theta$ by writing~\eqn{qbm3-b.13}
in terms of the squeeze parameter, with $f \equiv |f|e^{i\varepsilon}$:
\bal{3.613.7}       
\dot r &=& |f| \sin (2\phi + \varepsilon) \nn
\dot \phi &=& -h + |f|\coth 2r\cos (2\phi + \varepsilon) \nn
\dot \theta &=& h - |f|\th  r \cos (2\phi + \varepsilon)
\ea
These equations are useful for numerical work.
They also tell us of the existence of constant, and so possibly attractor,
solutions for $\phi, \theta$.  If we set $r\to\infty$ then the equations for
$\phi, \theta$ become
\bel{3.613.20-2}
\dot\theta = -\dot\phi = h-|f|\cos(2\phi + \varepsilon)
\ee
\begin{enumerate}
 \item Suppose there exist some $\theta$ and $\phi$ such that $\dot \theta =
   \dot\phi = 0$.  Then $h=|f|\cos(2\phi+\varepsilon)$, so that $|h|\le|f|$.  
   Thus, since $h$ is real, we have $h^2 \le |f|^2$, and from~\eqn{ent-4-4}
   this inequality is
   true if and only if $\Omega^2 \le 0$.
 \item Conversely suppose $\Omega^2 \le 0$.  
   Then by the previous argument,
   $|h|\le|f|$, or $-1 \le h/|f| \le 1$.  Thus there must exist some $\phi$ 
   such that $\cos(2\phi+\varepsilon)=h/|f|$.  From \eqn{3.613.20-2} we see
   that for this value of $\phi$, $\dot\theta = \dot\phi=0$.  
\end{enumerate}
In other words, there will exist constant solutions for $\phi, \theta$ if and
only if $\Omega^2\le 0$ (the oscillator is ``inverted'').  Of course, this 
doesn't reveal whether these constant
solutions are attractors.  Numerically solving \eqn{3.613.7} with 
$\Omega^2 \le 0$, for various $\cal E$, $\Omega$ 
and $\kappa$, shows that $\phi, \theta$ apparently do always quickly tend 
toward constants, always accompanied by one of $r\to\pm\infty$.

We note that it's common to eliminate the cross term in the
action by adding a surface term:
\bal{3.614-5}
L &\to& {M\over 2}\left(\dot x^2 + 2{\cal E}\dot xx - \Omega^2 x^2\right)
-{1\over 2} {d\over dt}(M{\cal E}x^2) \nn
&=& {M\over 2}\left[\dot x^2 - \left(\Omega^2+{\dot M{\cal E}\over M} + 
\dot{\cal E}\right) x^2\right]
\ea
Although this leaves the classical equation of motion unchanged, it will change
the squeeze parameters. In this paper we leave the cross term in our
lagrangians.

\section{Open Systems}

\subsection{Influence functional theory}
\Label{i-f-theory}

The influence functional (IF) formalism was first introduced by Feynman and
Vernon~\cite{if} as a way of deducing the influence of an
environment on some system of interest.  It was later applied by Caldeira
and Leggett~\cite{if} to the high temperature limit of a model 
where both system and environment are composed of static oscillators, that
is, having time independent frequency.
A comprehensive review is given by Grabert et al (in \cite{if}).

In these earlier works, the influence functional for quantum Brownian motion
has only been derived for Markovian processes corresponding to coupling to a
high temperature ohmic bath. An exact master equation for non-Markovian 
processes is recently derived by Hu Paz and Zhang \cite{HPZ1,HPZ2}
(see also \cite{Paz,HalYu}).
Hu and Matacz \cite{HM2} obtained the master equation for system and bath
oscillators with time-dependent frequencies, a result readily generalizable
to quantum fields. Stochastic properties of interacting quantum field theory
are discussed in \cite{Belgium,ZhangPhD}.
Most work in this area since Feynman and Vernon has assumed a bilinear
system-bath coupling, which yields an exact analytic form for the influence
functional. Recently, weak nonlinear couplings~\cite{HPZ2} have also been
considered using perturbation theory borrowed from field theory.

The language of influence functionals was developed in the context of
non-equilibrium statistical mechanics, but can be generalized to
field theory (see e.g., \cite{ZhangPhD}). In fact
it can be shown \cite{CH94} to be formally equivalent to the
Schwinger-Keldysh closed time path (CTP) formalism~\cite{ctp}.
Stochastic field theory based on the IF and CTP has since been applied to
semiclassical gravity \cite{Banff} and inflationary cosmology problems
\cite{CH95}.

In this paper we further develop the work of \cite{HM2} by considering
a squeezed system coupled bilinearly to a static bath (oscillators with
time-independent frequencies), but with a time-dependent coupling constant.
We also lay out the groundwork for calculating such
quantities as entropy and uncertainty as well as fluctuations and coherence,
for the purpose of this paper, and a later one on the de Sitter 
universe~\cite{KokMat}.

\subsection{Propagator for the density matrix}

The primary object we wish to consider is the evolution of the reduced
density matrix of our system via the Feynman-Vernon influence functional method.
This has been discussed at length
in~\cite{HM2}; we describe it here in order to establish the notation, and
just state its main results without deriving them.

Again consider our system described by $x$ which interacts with its environment
$q$ through some interaction.  The combined action is
\be
S[x,q] = S[x] + S_E[q] + S_{\it int}[x,q]
\ee
We require the reduced density matrix of the system at time $t$.  This is
found by tracing out the environment:
\be
\rho_r(x\,x'\,t) = \intinf dq\; \rho(x\,q\,x'\,q\,t)
\ee
The full density matrix $\rho(x\,q\,x'\,q\,t)$ evolves unitarily.  
Suppose we expand it using completeness relations and then path integrals:
\ba
\rho(x\,q\,x'\,q\,t) \!\!\!\!&=& \!\!\!\!\braket{x\,q\,t|\rho|x'\,q\,t} \nn
&=& \!\!\!\!\int dx_i\,dq_i\int dx_i'\,dq_i'\;\braket{x\,q\,t|x_i\,q_i\,0}
\braket{x_i\,q_i\,0|\rho|x_i'\,q_i'\,0}\braket{x_i'\,q_i'\,0|x'\,q\,t} \nn
&=&\!\!\!\!\int dx_i\,dq_i\int dx_i'\,dq_i'\int_{x_i}^xDx\int_{q_i}^q Dq\;
e^{iS[x,q]}\rho(x_i\,q_i\,x_i'\,q_i'\,0)\int_{x_i'}^{x'}Dx'\int_{q_i'}^q Dq'\;
e^{-iS[x',q']} \nn
&\equiv& \!\!\!\!\int dx_i\,dq_i\int dx_i'\,dq_i'\;
J(x\,q\,x'\,q\,t|x_i\,q_i\,x_i'\,q_i'\,0)\;\rho(x_i\,q_i\,x_i'\,q_i'\,0)
\ea
where $J$ is seen to be an evolution operator for the entire system plus
bath.  Now to allow further calculation we make the assumption that the system 
and bath are initially uncorrelated, i.e.
\be
\rho(x_i\,q_i\,x_i'\,q_i'\,0) = \rho_{\it sys}(x_i\,x_i'\,0)\;
\rho_E(q_i\,q_i'\,0)
\ee
(Initial conditions with correlations have also been considered by
\cite{HakAmb}).
In this case we are able to rearrange the order of integration to write
the reduced density matrix in the following way:
\bel{qbm3-2.5}
\rho_r(x\,x'\,t) = \int dx_i\,dx_i'\;J_r(x\,x'\,t|x_i\,x_i'\,0)\;
\rho_{\it sys}(x_i\,x_i'\,0)
\ee
where the evolution operator for the reduced density matrix is defined by
\be
J_r(x\,x'\,t|x_i\,x_i'\,0) \equiv 
\int_{x_i}^x\!Dx\!\int_{x_i'}^{x'}\!Dx'\,e^{iS[x]-iS[x']}\;F[x,x']
\ee
and $F[x,x']$ is the so-called influence functional:
\be
F[x,x'] = \int dq\;dq_i\;dq_i'\;\rho_E(q_i\,q_i'\,0)\;
\int_{q_i}^qDq\;e^{iS_E[q]+iS_{\it int}[x,q]}\;
\int_{q_i'}^qDq'\;e^{-iS_E[q']-iS_{\it int}[x',q']}
\ee
We can also write the influence functional in a basis-independent form 
as follows.  First we write the path integrals as propagators
\be
F[x,x'] = \int dq\;dq_i\;dq_i'\;\rho_E(q_i\,q_i'\,0)\;\braket{q|U(t)|q_i}\,
\braket{q_i'|U'^\dagger(t)|q}
\ee
where $U(t), U'(t)$ are the propagators for $S_E[q]+S_{\it int}[x,q]$ and
$S_E[q]+S_{\it int}[x',q]$ respectively.  Then upon integrating over $q, q_i$
and writing the remaining integral as a trace, we obtain:
\be
F[x,x'] = \tr U(t)\,\rho_E(0)\,U'^\dagger(t)
\ee
Using this form to calculate the influence functional was done earlier
in~\cite{HM2}. Here we just list the result: if we use
sum and difference coordinates defined by
\bel{qbm3-2.13}
\Sigma \equiv (x + x')/2 \;\;\;,\;\;\;
\Delta \equiv x - x'
\ee
then the influence functional can be written in terms of two new quantities, 
the ``dissipation'' $\mu(s,s')$ and ``noise'' $\nu(s,s')$:
\bel{hm2-2.8-3.1}
F[x,x'] = \exp{-1\over\hbar}\int_0^t ds\int_0^s ds'\;\Delta(s)
\Bigl[\nu(s,s')\,\Delta(s') + i\mu(s,s')\,2\Sigma(s')\Bigr]
\ee 
Thus the influence of the environment is completely invested in the dissipation
and noise.

\subsection{Evolution of the reduced density matrix}

Suppose now that we work within the context of quantum brownian motion, using 
the notation of~\cite{HM2}.  That is, our system is 
modeled by an oscillator with time dependent mass, cross term and natural 
frequency.  This interacts bilinearly with an environment modeled in the same
way, the total lagrangian being
\bal{qbm3-2.1}
S[x,\bold{q}] &=& S[x]+S_E[\bold{q}]+S_{\rm int}[x,\bold{q}] \nn
&& \nn
&=& \int_{t_i}^tds \left\{ {M(s)\over 2}\left( \dot x^2+ 2{\cal E}(s)x\dot{x}-
\Omega^2(s)x^2\right)\right. \nn
&&{} +\left.\sum_n\left[{m_n(s)\over 2}
  \left(\dot q_n^2+ 2 \varepsilon_n(s)q_n\dot q_n- \omega^2_n(s)q_n^2\right) 
\right]
  + \sum_n\left[-c(s)xq_n\right]\right\}   
\ea
where the particle and the bath oscillators have coordinates $x$ and $q_n$
respectively.

We wish to start with some initial system density matrix $\rho_{\it sys}
(x_i\,x_i'\,0)$ and evolve it using~\eqn{qbm3-2.5}.
As described in~\cite{HM2}, $J_r$ is calculated using the standard
path integral approach. 
Using the sum and difference coordinates defined in~\eqn{qbm3-2.13}, the 
classical paths followed by the system, $\Sigma_{\it cl}, \Delta_{\it cl}$,
can be written in terms of more elementary functions $u,v$:
\bal{qbm3-3.9}
\Sigma_{\it cl}(s) &=& \Sigma_{\it cl}(t_i)u_1(s) + \Sigma_{\it cl}(t)u_2(s)\nn
\Delta_{\it cl}(s) &=& \Delta_{\it cl}(t_i)v_1(s) + \Delta_{\it cl}(t)v_2(s)
\ea
Then it can be shown that the superpropagator $J_r$ is equal to
\bal{qbm3-3.13}
J_r(x,x',t|x_i,x'_i,t_i) &=& {|b_2|\over 2\pi\hbar}\exp\left[{i\over
\hbar}(b_1\Sigma\Delta-b_2\Sigma\Delta_i+b_3\Sigma_i\Delta-b_4\Sigma_i
\Delta_i)\right. \nn
&&{}-\left.{1\over\hbar}\left(a_{11}\Delta_i^2+a_{12}\Delta_i\Delta
+a_{22}\Delta^2\right)\right]
\ea
The functions $b_1\rightarrow b_4$ can be expressed as
\bal{qbm3-3.11}
b_1(t,t_i) &=& M(t)\dot u_2(t)+{M(t){\cal E}(t)} \nn
b_2(t,t_i) &=& M(t_i)\dot u_2(t_i) \nn
b_3(t,t_i) &=& M(t)\dot u_1(t) \nn
b_4(t,t_i) &=& M(t_i)\dot u_1(t_i)+{M(t_i){\cal E}(t_i)}
\ea
while the functions $a_{ij}$ are defined by
\bel{qbm3-3.12}
a_{ij}(t,t_i)={1\over 1+\delta_{ij}}\int_{t_i}^t ds\int_{t_i}^t ds'\;
v_i(s)\;\nu(s,s')\;v_j(s')
\ee
The functions $u_1\to v_2$ are solutions to the following equations (dropping
subscripts on $u,v$):
\bel{qbm3-3.5}
\ddot u(s)+{\dot M\over M}\dot u+
\left(\Omega^2+\dot {\cal E}+{\dot M \over M}{\cal E}\right)u+
{2\over M(s)} \int_{t_i}^s ds'\;\mu(s,s')\;u(s')=0
\ee
\bel{qbm3-3.6}
\ddot v(s)+{\dot M\over M}\dot v+
\left(\Omega^2+ \dot {\cal E}+ {\dot M\over M}{\cal E}\right)v-
{2\over M(s)} \int_s^t ds'\;\mu(s,s')\;v(s')=0
\ee
subject to the boundary conditions
$$
u_1(t_i) = v_1(t_i) = 1\;\;\;,\;\;\;
u_1(t)   = v_1(t)   = 0
$$
\be
u_2(t_i) = v_2(t_i) = 0\;\;\;,\;\;\;
u_2(t)   = v_2(t)   = 1
\ee

\subsection{Propagator $J_r\,$ for the reduced density matrix: ohmic 
environment}
\Label{superpropagator}

To proceed further we need explicit expressions for $a_{11} \rightarrow b_4$.  
These are expressed
in terms of $u_1 \rightarrow v_2$, which in turn come from 
solving~(\ref{qbm3-3.5}, \ref{qbm3-3.6}).  To solve these equations we need to
know the dissipation~$\mu$ of the environment.  

The noise and dissipation can be calculated from~\cite[eqns 2.18, 2.19]{HM2}.  
We choose the bath oscillators to be simple harmonic, that is, static with no 
cross term, since this turns out to correspond to the simplest form of
dissipation: local.  
For such an environment the dissipation and noise can be shown to be
\bal{hm2-2.18-2.19}
\mu(s,s') &=& \int_0^\infty d\omega\;I(\omega,s,s')\im [X(s)X^\ast(s')] \nn
\nu(s,s') &=& \int_0^\infty d\omega\;I(\omega,s,s') \coth {\omega\over 2T} \re
[X(s)X^\ast(s')]
\ea
where by $T$ we will always mean $k_BT/\hbar$; $X$ is the sum of the Bogoliubov
coefficients for the bath oscillators and $I$ is the ``spectral density'', 
a function defined by
\bel{qbm3-2.26}
I(\omega, s, s') = {c(s)c(s')\over 2\kappa}\sum_n \delta(\omega-\omega_n)
\ee
which encodes information of the action of the environment on the system.
In general the spectral density can be described by some function of 
$\omega^j$, where $j$ is set by the particular environment being modeled.
The case of $j=1$, a so-called ``ohmic'' environment, is a borderline between
the super-ohmic case ($j>1$)---which models weak damping---and the subohmic
case ($j<1$) modelling strong damping.  We can in effect consider both damping
extremes by taking an ohmic environment together with some strength 
$\gamma_0$ which can be altered from zero, for a free system, up to higher 
values.

Also, by considering the continuum limit of the coupling constant, it can be
shown that this constant's independence of $n$ also leads to an ohmic 
environment; so we will only consider spectral densities of the following form:
\bel{3.601.9-1}
I(\omega,s,s') = {2\gamma_0\over\pi}\, \omega\; c(s)c(s')
\ee
For a general lagrangian the sum of the Bogoliubov coefficients $X$ will be 
complicated; 
however we have simplified our calculations by taking the bath to 
be composed of unsqueezed (i.e.\ coherent) static oscillators with unit mass.  
For this type of bath the dissipation and noise can be calculated
for an
arbitrary bath temperature; we use the integral form of the noise as being
easier to work with:
\bal{mu-and-nu} 
\mu(s,s') &=& 2\gamma_0\,c(s)c(s')\,\delta'(s-s') \nn
\nu(s,s') &=& {2\gamma_0\over \pi}\,c(s)c(s')\int_0^\infty \omega\coth{\omega
\over 2T}\cos \omega(s-s')\;d\omega 
\ea
In the high temperature limit the noise becomes white, that is it tends toward
a delta function.

\section{Entropy and uncertainty, fluctuations and coherence}

\subsection{Initial and final states}

Assume the systems are initially in the vacuum state, so that
their density matrix is gaussian.  So we start with an arbitrary gaussian 
reduced density matrix
\bel{3.723-1}
\rho_r(x_i\;x_i'\;t_i) \propto e^{-\xi x_i^2 + \chi x_i x_i' -\xi^\ast x_i'^2}
\ee
and propagate it by using~(\ref{qbm3-2.5},~\ref{qbm3-3.13}) to give
\bel{3-0.5}
\rho_r(x,x',t)  = N e^{-A \Delta^2 - 2iB\Delta\Sigma-4C\Sigma^2}
\ee
where we have used the same $A$, $B$ and $C$ notation of~\cite{Mat},
and with $\xi_r, \xi_i$ the real and imaginary parts of $\xi$:
\bal{3.725}
N &=& 2\sqrt{C/\pi} \nn
A &=& a_{22}+{1\over D}\left\{[(2\xi_r+\chi)/4 + a_{11}]\,b_3^2 +(2\xi_i+b_4)\,
a_{12}\,b_3 -(2\xi_r-\chi)a_{12}^2\right\} \nn
B &=& -b_1/2+{1\over D}[(\xi_i+b_4/2)\,b_2\,b_3-(2\xi_r-\chi)a_{12}\,b_2]\nn
C &=& {1\over 4D}(2\xi_r-\chi)\,b_2^2 \nn
&&\nn
D &=& 4|\xi|^2-\chi^2+4\,(2\xi_r-\chi) a_{11} +4\,\xi_i \,b_4 + b_4^2 
\ea
These expressions form the basis of our later calculations.
The quantity we are focusing on is the reduced density
matrix,~\eqn{3-0.5}, using the expressions in~\eqn{3.725}.  These in turn
use~\eqn{3.870.1}, which depends on our obtaining $X$, the sum
of the Bogoliubov coefficients for the effective oscillator.

\subsection{Entropy from the reduced density matrix}
\Label{entropy-section}

The entropy of a field mode has been calculated by Joos and Zeh~\cite{JooZeh}. 
 It can be
derived from the reduced density matrix at time $t$
by using~\eqn{von-neumann-entropy}, and is given by
\bel{3-1}
S = {-1\over w} [w\ln w + (1-w) \ln (1-w)] \simeq 1-\ln w \;\;\;
\mbox{if $w\rightarrow 0$}
\ee
where 
\bel{3-2}
w \equiv {2\sqrt{C/A}\over 1+\sqrt{C/A}}
\ee
The linear entropy is often more useful to work with owing to its simplicity:
\bel{3-2.5}
S_{\it lin}\equiv -{\rm tr}\,\rho^2 = -\sqrt{C/A}
\ee
and $S=0\to\infty$ is equivalent to $S_{\it lin} = -1\to 0$, both strictly
increasing.  Then if $S_{\it lin} \to 0$ we have
\bel{ent-3-3}
S \to -\ln |S_{\it lin}| + 1-\ln 2 \;\;\;,\;\;\;
{\rm i.e.\ } S_{\it lin}\to -e^{1-S}/2
\ee
As an example, suppose we have a system in an initially pure gaussian state
($\chi = 0$), so that noise and dissipation are absent: $\gamma_0=0$.  In this 
case, from~(\ref{mu-and-nu}, \ref{3.870.1}) we have 
\be
a_{11}=a_{12}=a_{22} =0
\ee
so that \eqn{3.725} gives $C/A=1$ and hence from \eqn{3-1} $S =0$ as expected.

\subsection {Fluctuations and coherence}

A clearer picture of the dynamics of a closed and open system can be
obtained if we rotate the phase space axes so that
the density matrix can be expressed in terms of the so called super- and
subfluctuant variables.  (Alternatively, we are rotating the Wigner function
in phase space so as to eliminate the cross term there).  Call these variables
$u,v$, expressed as real linear 
combinations of $q,p$ [they have nothing to do with the $u, v$ 
of~\eqn{qbm3-3.9}].  We fix the linear combinations such that one variable
($u$, the superfluctuant) grows exponentially while the other decays 
exponentially.
In the case of no coupling to the environment we proceed by expressing
$\braket{u^2}, \braket{v^2}$ in terms of $\braket{q^2},\braket{qp+pq},
\braket{p^2}$, and then substituting for these the standard squeezed state 
results~\cite{Mat}.  This enables us to write
\bel{3.825.1-1.5-and-1.6}
\braket{u^2} = {\kappa e^{2r}\over 2}\;\;\;,\;\;\; \braket{v^2} = {e^{-2r}
\over 2\kappa}
\ee
These relations fix $u,v$ in terms of $q,p$, and we now use the same 
transformation for the case of nonzero dissipation:
\bal{3.825.1-2}
u &=& -\kappa\sin\phi\,q+\cos\phi\,p \nn
v &=& \cos\phi\,q +{\sin\phi\over\kappa}\,p
\ea
What we wish to do is take a density matrix in position, \eqn{3-0.5}, and 
write it in the $u,v$ basis.  Consider first of all calculating $\rho(u,u')$:
\bel{3.825.3-1}
\rho(u,u') = \int \braket{u|q}\,\rho(q,q')\,\braket{q'|u'}\;dq\,dq'
\ee
We need $\braket{u|q}$.  This can be found by solving the p.d.e which follows
by quantising~\eqn{3.825.1-2} and applying both sides to $\braket{q|u}$:
\be
u\braket{q|u} = \left(-\kappa\sin\phi \;q-i\cos\phi\;\partial_{q}\right)
\braket{q|u}
\ee
which has solution
\be
\braket{q|u} = f(u) \exp{i\over\cos\phi}
\left[{\kappa\sin\phi \,q^2\over 2}+q u
\right]
\ee
for some function $f(u)$ to be determined [unrelated to~\eqn{ent-4-4}].  
We determine $f(u)$ by redoing this calculation
with the roles of $q$ and $u$ interchanged; since $[v,u] = i$, we have
\be
q\braket{u|q} = \left({-\sin\phi\, u\over\kappa} + i\cos\phi\partial_u\right)
\braket{u|q}
\ee
Solving this determines $f(u)$ and allows us to finally write (up to a phase)
\bel{3.825.5-1}
\braket{q|u} = {1\over\sqrt{2\pi\cos\phi}}\exp{i\over\cos\phi}
\left[{\kappa\sin\phi \,q^2\over 2}+q u+{\sin\phi \,u^2\over 2\kappa}\right]
\ee
Similarly we find
\bel{3.825.5-4}
\braket{q|v} = \sqrt{\kappa\over 2\pi\sin\phi}\exp{i\kappa\over\sin\phi}
\left[{-\cos\phi \,q^2\over 2}+q v-{\cos\phi \,v^2\over 2}\right]
\ee
Now, suppose we start with a gaussian density matrix as in~\eqn{3-0.5}.
We can then easily change bases using (\ref{3.825.3-1}, \ref{3.825.5-1}, 
\ref{3.825.5-4}) to get, with
\bel{3.830-2-and-3.830-1}
\gamma \equiv {\kappa\over 2}\cot\phi \;\;\;,\;\;\;
\sigma \equiv {\sin^2\phi\over \kappa^2}\left[4AC+(B-\gamma)^2\right]
\ee
\bel{3.830.1-1}
\lambda \equiv {4AC+(4\gamma\sigma+B-\gamma)^2\over 4\sigma^2}
\ee
\bal{3.830-3}
\rho(u,u') &=& \sqrt{C\over\pi\sigma\lambda}\exp{-1\over 4\sigma\lambda}
\left[A\Delta_u^2+2i(4\gamma\sigma+B-\gamma)\Delta_u\Sigma_u+4C\Sigma_u^2
\right] \nn
&&\nn
\rho(v,v') &=& \sqrt{C\over\pi\sigma}\exp{-1\over 4\sigma}\left[A\Delta_v^2
-2i(4\gamma\sigma+B-\gamma)\Delta_u\Sigma_u+4C\Sigma_v^2\right]
\ea
where we have used sum and difference variables, e.g.\ $\Sigma_u \equiv 
(u+u')/2, \;\Delta_u \equiv u-u'$, and $\gamma$ has no relation to $\gamma_0$.

We can show that in the absence of a bath, these matrices reduce to the 
expected ones for a squeezed vacuum.  First, in the 
$q$-representation the density matrix of a squeezed vacuum is known to be
\cite{HYV}
\bel{coh-state-3.37}
\rho(q,q') \propto {-\kappa\over 2}{1+e^{2i\phi}\tanh r \over 1-e^{2i\phi}
\tanh r}\left(q^2+q'^2\right)
\ee
If we write $\rho(q,q')$ in terms of sum and difference coordinates and 
compare with the definitions of $A, B, C$ in~\eqn{3-0.5}, we find
\ba
A = C &=& {\kappa\over 4} \;{1-\tanh^2r\over 1-2\cos 2\phi\tanh r+\tanh^2 r} 
\nn
&&\nn
B &=& {\kappa\sin 2\phi \tanh r\over 1-2\cos 2\phi\tanh r+\tanh^2 r}
\ea
Substituting these into \eqn{3.830-3} gives
\bal{3.830.4-and-5}
\rho(u,u') &=& {e^{-r}\over\sqrt{\pi\kappa}} \exp{-e^{-2r}\over 2\kappa}
\left(u^2+u'^2\right) \nn
\rho(v,v') &=& \sqrt{\kappa\over\pi}e^r\exp{-\kappa e^{2r}\over 2}\left
(v^2+v'^2\right)
\ea
These are the expected results, as can be seen by the fact that with
$p, q$ replaced by $u, v$ respectively, they are produced when $\phi$ is
set to zero in $\rho(p,p')$ and $\rho(q,q')$.

\subsubsection*{Measures of fluctuations and coherence}

Returning to the general case of dissipation, the fluctuations in $u$ and $v$ 
are calculated from the density matrices:
\bal{3.830.1-4}
\Delta u^2 &=& \braket{u^2} -\braket{u}^2 = \int u^2 \rho(u,u)\;du -
\left[\,\int u \,\rho(u,u)\;du\right]^2 = {\sigma\lambda\over 2C} \nn
\Delta v^2 &=& {\sigma\over 2C}
\ea
and both of these are just equal to 1/2 divided by the coefficient of 
$-\Sigma^2$ in their density matrix.

As a measure of coherence we note that a large 
coefficient of $-\Delta^2$ means that the density matrix
is strongly peaked along its diagonal, i.e.\ there is very little 
coherence in the system. A measure of coherence was 
defined in~\cite{Mat93} as a squared coherence length $L^2$, equal 
to 1/8 divided by the coefficient of $-\Delta^2$, so that
a large $L^2$ means a high degree of coherence in the 
system.  With this definition of $L^2$, \eqn{3.830-3} gives 
\bel{3.830.1-2}
L_u^2=\frac{\sigma\lambda}{2A},\;\;\;L_v^2=\frac{\sigma}{2A}
\ee
We can also relate the coherence lengths and fluctuations to the entropy
of the system (see section~\ref{entropy-section} for definitions).  We can 
write
\bel{3.830.1-5}
{L_u^2\over\Delta u^2} = {L_v^2\over\Delta v^2}=S_{\it lin}^2 = {C\over A}
\ee
(A note of caution: linear entropy is negative by definition in order for it
to increase with $S$.  Then as $S_{\it lin}$ increases, $S_{\it lin}^2$ will
decrease).  Also the uncertainty relation for $u, v$ becomes, 
from~\eqnnn{3.830-2-and-3.830-1}{3.830.1-1}{3.830.1-4}:
\bel{dec-3.830.2-6}
\Delta u^2 \Delta v^2 = {1\over S_{\it lin}^2}\left[{1\over 4} +
{(4\gamma\sigma+B-\gamma)^2\over 16AC}\right]
\ee
For the free field the last term in the square brackets is zero while 
$S_{\it lin} = -1$ (since $S=0$), so that $\Delta u\Delta v = 1/2$.

\section{Entropy and uncertainty of oscillator system}

We can now demonstrate how the previous results are used. In the simplest cases,
such as a static oscillator coupled to a thermal bath of static oscillators,
with a static ohmic coupling, the entropy is easily
compared with known results in equilibrium statistical mechanics.
{}From section~\ref{superpropagator}, we know that this case has
local dissipation [i.e.\ $\mu \propto \delta'(\Delta)$], and at
$T\to\infty$  the noise becomes white [$\nu \propto \delta(\Delta)$].

\begin{figure}[htb]
\epsfxsize=10cm
\epsfysize=7cm
\centering{\ \epsfbox{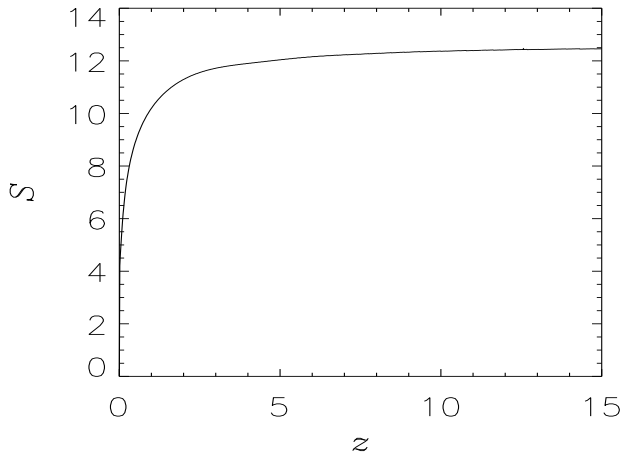}}
\vspace{-0.5cm}
\caption{Entropy growth over time.\label{3.613.3.eps}}
\end{figure}

For thermal equilibrium, the standard statistical mechanics result for the
entropy at high temperature is
\bel{2.05-5}
S \rightarrow 1+\ln{T\over k}
\ee
Obtaining this result with this formalism is a good example in its
application. We will leave the details in Appendix B but show the numerical
results in plots.
Figure~\ref{3.613.3.eps} shows a plot of
$S$-vs-$z$ for $\sigma=1, k = 1, \gamma_0=0.1, T = 10^5$.
For these numbers, \eqn{2.05-5} gives $S \to 12.513$ as $z\to\infty$,
as compared with $S \to 12.514$ numerically at $z=100$, a result indicated
by the figure. The relaxation time, defined to be
\bel{3.613.3-2}
{1\over 2\gamma_0}=5
\ee
is apparent in the figure as a characteristic time over which the entropy
climbs to its final value, while the decoherence time scale
\cite{PHZ}
\bel{3.613.3-1}
{1\over 4M\gamma_0T\sigma^2} = 2.5\times 10^{-5}
\ee
is too small to be noticeable.

\subsubsection*{Coherent state as the state of least entropy}

We now use our entropy expression to investigate the claim that for 
large times the state of least entropy for the static oscillator is the 
coherent one, at least for white noise and local dissipation.  This was shown 
in~\cite{ZHP} in the small $\gamma_0$ limit by using a Wigner function
approach.  

\begin{figure}[htb]
\epsfysize=6cm
\centering{\ \epsfbox{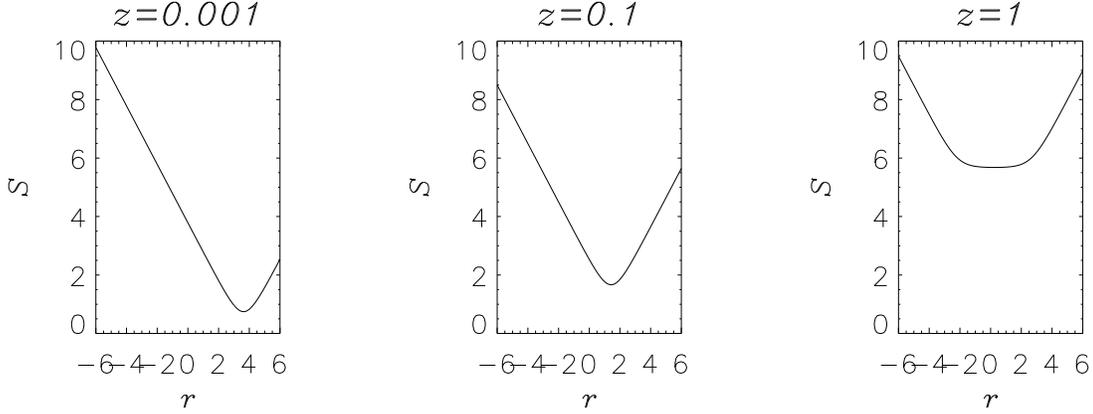}}
\vspace{-0.5cm}
\caption{Entropy at various times.\label{3.613.2.eps}}
\end{figure}

Using our expression for the entropy $S$, we can plot $S$ versus the initial
squeeze parameter $r$ for various times in figure~\ref{3.613.2.eps}.
We have chosen $k=10, \gamma_0 = 0.1$.  The squeeze
parameter~$r$ is related to~$\sigma$, the width of the gaussian wavefunction, by
\be
r \equiv \ln {\sigma_0\over\sigma} \;\;\; ; \;\;\;
\sigma_0 \equiv \sqrt{1\over 2\kappa}
\ee
or,
\be
\sigma = {e^{-r}\over \sqrt{2\kappa}}
\ee
Note that at early times (e.g.\ $z=0.001$), the entropy is minimised for high
initial squeezing, as noted in~\cite[fig.\ 1]{ZHP}; this is not unreasonable
since such a highly squeezed state will spread with time, becoming 
indistinguishable at later times from states which started out being less
highly squeezed.  At late times the 
entropy is minimised by starting with small or zero 
squeezing, i.e.\ an initially coherent state is the one which minimises entropy
at late times.  Thus our approach agrees with~\cite{ZHP}, and may be more 
useful in that it allows us to directly calculate the entropy at all times.

\subsection{Static inverted oscillator}

The static inverted oscillator is the simplest squeezed system.
It also models the zero mode of the inflaton field in New Inflation \cite{infcos}.
Its lagrangian is:
\bel{3.800-1}
L(t) = {1\over 2}[\dot x^2 + k^2 x^2]
\ee
Suppose this is coupled to the usual environment of harmonic oscillators in
a thermal state, with coupling constant $c(s)=1$.
Then the equivalent oscillator we consider has unit mass, no cross term and 
frequency 
\be
\Omega_{\it eff}^2 = -k^2-\gamma_0^2 \equiv -\kappa^2
\ee
so that from \eqn{5-1} the sum of its Bogoliubov coefficients is 
(taking $t_i=0$)
\be
X(t) = \ch  z -i\sh z
\ee
Hence from~\eqn{3.752.3.1-1} we have
\bel{3.870.9-1}
\alpha = \ch z \;\;\;,\;\;\; \beta = -i\sh z
\ee
so that from~\eqn{3.870.8-4.5} at late times ($z\to\infty$)
\bel{3.870.9-4}
r \to z
\ee
To investigate the dependence of the entropy on the various quantities in the
propagator coefficients, we calculate these coefficients first for white noise
analytically; we then calculate them numerically for zero temperature. 

The $b_i$'s are independent of the temperature, and using~\eqn{3.870.1} they 
are found to be (where here and elsewhere a carat will denote division by 
$\kappa$)
\bel{3.801-4}
b_{\{^1_4\}} =  \kappa (\pm\coth z - \hat\gamma_0 ) \;\;\;,\;\;\;
b_{\{^2_3\}} = {\pm \kappa e^{\pm \hat\gamma_0 z}\over \sh z}
\ee

\subsubsection*{High temperature}

White noise is given by $\nu(s,s') = 4\gamma_0 T\;\delta(s-s')$, or
$\nu(\zeta,\zeta') = 4 \hat\gamma_0   \kappa^2 T \delta(\zeta-\zeta')$; 
the relevant quantities are inserted into~\eqn{3.870.1} with the $a_{ij}$'s 
then becoming
\bal{3.802}
a_{11} &=& {T\over 2\hat k^2\shsq z} \left[\hat k^2 +
e^{2\hat\gamma_0 z}-\hat\gamma_0 \sh 2z-\hat\gamma_0 ^2 \ch 2 z\right] \nn
&& \nn
a_{12} &=& {Te^{-\hat\gamma_0 z}\over \hat k^2\shsq z} \left[
\left(1-e^{2\hat\gamma_0 z}\right)\ch z+
\left(1+e^{2\hat\gamma_0 z}\right)\hat\gamma_0 \,\sh z 
\right] \nn
&& \nn
a_{22} &=& {Te^{-2\hat\gamma_0 z}\over 2\hat k^2\shsq z} \left[-\hat k^2 
e^{2\hat\gamma_0 z}-1 +\hat\gamma_0  e^{2\hat\gamma_0 z}
\left(\hat\gamma_0 \, \ch 2 z -\sh 2z\right)\right]
\ea
Note that $\hat\gamma_0=\gamma_0/\kappa < 1$; however if we assume small 
dissipation ($\hat\gamma_0 \ll 1$) we can write down large 
time limits of these quantities:
$$
a_{11} \to {T\hat\gamma_0 \over 1-\hat\gamma_0 } \;\;\;,\;\;\;
a_{12} \to {2Te^{-(1-\hat\gamma_0 )z}\over 1+\hat\gamma_0 } \;\;\;,\;\;\;
a_{22} \to {T\hat\gamma_0 \over 1+\hat\gamma_0 }
$$
\bel{3.802.1}
b_{\{^1_4\}} \to  \kappa(\pm1 - \hat\gamma_0 ) \;\;\;,\;\;\;
b_{\{^2_3\}} \to \pm 2 \kappa \,e^{-(1\mp \hat\gamma_0 )z}
\ee
We can now calculate large time limits of the density matrix coefficients 
from~\eqn{3.725}:
\bel{3.840.2-4}
A \to a_{22} \;\;\;,\;\;\;
B \to -b_1/2 \;\;\;,\;\;\;
C \to {b_2^2\over 16 a_{11}}
\ee
These coefficients are independent of the initial conditions, which might be
expected since the dissipation is acting to damp out any late time dependence 
on these initial conditions.  So we have
\bel{3.844-5}
S_{\it lin} = -\sqrt{C \over A} \to {-\kappa^2 e^{-z} \over 2\gamma_0 T}
\ee
so that from (\ref{ent-3-3}, \ref{3.870.9-4})
\bel{3.844.1-1}
S \to r + 1+\ln {T\gamma_0\over  \kappa^2}
\ee

\subsubsection*{Zero temperature}

At $T=0$, the action of the environment is due to quantum effects
only.  
If we write the noise in its 
primitive form as the usual integral over frequency then we can leave this
frequency integration until last after the time integrations have been done.  
We will follow a more sophisticated approach in
a later paper~\cite{KokMat}, but we show it here to investigate what value
it might have.

So we refer to~(\ref{3.870.1}, \ref{mu-and-nu}), swapping the limits of 
integration to write
\bal{3.840.8-2-and-6}
a_{11} &=& {\gamma_0 \over \pi\shsq z}\int_0^{\hat{\omega}_{\rm max}} 
d\hat\omega\;\hat\omega\coth {\hat\omega \kappa\over 2T} 
\int_0^z d\zeta\int_0^z d\zeta' e^{\hat\gamma_0 (\zeta+\zeta')}\sh (z-\zeta)
\sh (z-\zeta')\cos \hat\omega(\zeta-\zeta') \nn
&&\nn
&=& {\gamma_0 \over 2\pi\shsq z}
\int_0^{\hat{\omega}_{\rm max}} d\hat\omega\;\hat\omega\coth {\hat\omega \kappa
\over 2T} \;I_{11}
\ea
where 
\bal{3.840.8-5}
I_{11} &\equiv& \left\{\hat k^2-\hat\omega^2+2 e^{2\hat\gamma_0 z}+\left(
1+\hat\gamma_0^2+\hat\omega^2\right)\ch 2z \right. \nn
&&{}-\left.4 e^{\hat\gamma_0 z}[\cos \hat\omega z\;(\ch z+\hat\gamma_0 \,
\sh z)+\hat\omega\sin \hat\omega z\sh z] + 2\hat\gamma_0 \,\sh 2z\right\}
\Big/ \nn
&&\left[\hat k^4 +2\hat\omega^2 \left(1+\hat\gamma_0^2\right)+\hat\omega^4
\right]
\ea
Similarly
\bel{3.840.9-2}
a_{12} = {\gamma_0e^{-\hat\gamma_0z}\over \pi\shsq z}\int_0^{\hat
\omega_{\rm max}} d\hat\omega\;\hat\omega\coth {\hat\omega \kappa\over 2T}
\;I_{12}
\ee
where 
\bal{3.840.9-1}
I_{12} &\equiv& \left\{-2\ch z\left(1+e^{2\hat\gamma_0z}\right)-
2\hat\gamma_0\sh z\left(1-e^{2\hat\gamma_0z}\right)\right.\nn
&&{}+\left.e^{\hat\gamma_0z}\cos \hat\omega z\left[3+\hat\gamma_0^2+\hat
\omega^2 +\left(\hat k^2-\hat\omega^2\right)\ch 2z\right]
+ 2\hat\omega e^{\hat\gamma_0z}\sin \hat\omega z \sh 2z\right\}\Big/ \nn
&&\left[\hat k^4 +2\hat\omega^2 \left(1+\hat\gamma_0^2\right)+\hat\omega^4
\right]
\ea
and
\bel{3.840.9-4}
a_{22} = {\gamma_0e^{-2\hat\gamma_0z}\over 2\pi\shsq z}
\int_0^{\hat\omega_{\rm max}} d\hat\omega\;\hat\omega\coth {\hat\omega \kappa
\over2T} \;I_{22}
\ee
where 
\bal{3.840.9-3}
I_{22} &\equiv& \left\{2 + e^{2\hat\gamma_0z}\left[\hat k^2-\hat\omega^2+
\left(1+\hat\gamma_0^2+\hat\omega^2\right)\ch 2z - 2\hat\gamma_0\sh 2z
\right]\right.\nn
&&{}+\left.4e^{\hat\gamma_0z}\left[\cos \hat\omega z\left(-\ch z + \hat
\gamma_0\sh z\right) - \hat\omega\sin \hat\omega z \sh z\right]\right\}
\Big/ \nn
&&\left[\hat k^4 +2\hat\omega^2 \left(1+\hat\gamma_0^2\right)+\hat\omega^4
\right]
\ea
With $T=0$ the coth term is set to one.  Then in all cases $a_{ij}$ starts at 
zero at $z=0$; for low dissipation $a_{11}, 
a_{22}$ quickly climb to similar constant values while $a_{12}$ climbs briefly 
but then rapidly decreases to zero.  This behaviour quantitatively matches the 
large time limits of the white noise $a_{ij}$'s in~\eqn{3.802.1}, even though 
the two calculations were done quite differently.  The asymptotic value of 
$a_{11}$ increases in even steps as we increase $\hat\omega_{\rm max}$
exponentially.  So we can make $a_{11}$ arbitrarily large by taking a large 
enough cutoff, so that it will always dominate $D$.

In that case, with $\hat\gamma_0 \ll 1$ we have at late times, using the
$b_i$'s in~\eqn{3.802.1}
\bel{3.840.2.1-1}
A \to a_{22} \;\;\;,\;\;\;
B \to -b_1/2 \;\;\;,\;\;\;
C \to {b_2^2\over 16a_{11}}
\ee
Again the coefficients are independent of the initial conditions.
Since $b_2$ is unchanged from the high temperature case and $a_{11}, a_{22}$
tend toward constants, we now can say
\bel{3.844.1-2}
S_{\it lin} \to {-\kappa e^{-z} \over 2\sqrt{a_{11}a_{22}}}
\ee
and so again from (\ref{ent-3-3}, \ref{3.870.9-4})
\be
S \to r + 1+\ln{\sqrt{a_{11}a_{22}}\over \kappa}
\ee

\section{Scalar field in de Sitter spacetime}
\Label{de-sitter-section}

We now turn to an example in cosmology, that of an inflationary universe
\cite{infcos}. We want to calculate the entropy of
a massless scalar field minimally coupled to gravity in a de~Sitter spacetime
by examining the evolution of the density matrix.  As we shall see, it is
a generally solvable squeezed system.

Consider  a scalar field $\Phi$ of mass $m$, described by the lagrangian density
\bel{3.631-5.1}
{\cal L} = {\sqrt{-g}\over 2}\left[g^{\mu\nu}\Phi,_{\mu}\Phi,_{\nu}
-\left(m^2+\xi R\right)\Phi^2\right]
\ee
coupled  by $\xi$ to the curvature $R=6(\dot a^2/a^2 + \ddot a/a$) of
a spatially flat FRW universe with metric
\be
ds^2 = dt^2 - a^2(t)\sum_i (dx^i)^2
\ee
In conformal time $\eta = \int dt/a$, the conformally-related field
$\chi = a \Phi$ is described by a Lagrangian density
\bel{3.631-2}
{\cal L} = {1\over 2}\left[\chi'^2-\sum \chi,_i^2-2{a'\over a}\chi\chi'
-\chi^2\left(m^2a^2-{a'^2\over a^2}+6\xi{a''\over a}\right)\right]
\ee
Decomposing the field into normal modes $k$ with amplitudes $q_k$, the
Lagrangian can be expressed as
\bel{3.631.5-1}
L(\eta) = \sum {1\over 2}\left[q'^2-2{a'\over a}qq' - 
q^2 \left(k^2+m^2a^2-{a'^2\over a^2}+6\xi{a''\over a}\right)\right]
\ee
Inside the brackets if we  add a surface term of $6\xi (q^2a'/a)'$ to
eliminate the $a''$ term (for justification, see \cite{HM2}), we
get a  new lagrangian:
\bel{3.631.5-5}
L_{\it new}(\eta) = \sum {1\over 2}\left[q'^2+2(6\xi-1){a'\over a}qq' - 
q^2 \left(k^2+m^2a^2+(6\xi-1){a'^2\over a^2}\right)\right]
\ee
For a massless minimally coupled scalar field in de Sitter space,
\bel{3.631.8-5}
L_{\it new}(\eta) = \sum{1\over 2}\left[q'^2+{2\over\eta}qq'-q^2\left(k^2-
{1\over\eta^2}\right)\right]
\ee
We also use a spectral density of the form
\bel{3.855-1}
I(\omega, \eta, \eta') = {2\gamma_0\over\pi H}{\omega\over\sqrt{\eta\eta'}}
\ee
so that $c(\eta) = 1/\sqrt{-H\eta}$.  This form of spectral density will be
justified in a later paper~\cite{KokMat}, although for now we note that it does
not make the equation of motion for $X$ any harder to solve than if we had
used a static coupling.  Since $\gamma_0/H$ is dimensionless we 
rewrite it as $c$ [not to be confused with $c(\eta)$].
Incorporating the bath gives the equivalent oscillator with $M=1, {\cal E} = 
1/\eta$ and frequency, from~\eqn{3.601.10-2},
\bel{3.870.5-0.1}
\Omega_{\it eff}^2 = k^2 - {1+c^2\over \eta^2}
\ee
Also we choose $\kappa = k$ to simplify the equation of motion.
With $z = k\eta$ we can write this together with its initial conditions 
from~(\ref{4-3}, \ref{5-1}, \ref{3.751}) as
$$
X''(z) +\left(1-{2+c^2\over z^2}\right)X =0
$$
\bel{3.870.5-1}
X(z_i) = 1 \;\;\;,\;\;\;
X'(z_i) =  -i-1/z_i
\ee
where $z<0$.
The solution of this equation can be constructed using Bessel functions whose
index is a function of $c$; however since we are interested in small 
$c$ we take the solution to be approximately that of the same equation
but with $c$ set to zero.  This simplifies things greatly:
\bel{3.870.2.1-1}
X(z) = \left(1+{i\over 2z_i}\right)f(z)+{i\over z_i}f^\ast (z)
\ee
where 
\be
f(z) \equiv \left(1-{i\over z}\right)e^{i(z_i-z)}
\ee
We can further simplify $X$ by using a very early initial time, setting 
$z_i \to -\infty$.
We also disregard the phase in the resulting expression for $X$, since this is 
not expected to make any difference to physical quantities.  In this case we 
obtain a new function which we rename $X$:
\bel{3.870.2.1-3}
X(z) \leadsto \left(1-{i\over z}\right)e^{-iz}
\ee
The Bogoliubov coefficients can now be found from \eqn{3.752.3.1-1}:
\bel{3.870.8-3}
\alpha = \left(1-{i\over 2z}\right)e^{-iz} \;\;\;,\;\;\;
\beta = {-i\over 2z}e^{-iz}
\ee
and so from \eqn{3.870.8-4.5} at late times
\bel{3.870.8-5}
r \to -\ln|z|
\ee
This result was also obtained in~\cite{Mat} using a different
formalism.

First we calculate the $b_{i}$'s.  Since we are only interested in late times
we can work to leading order in~$z$ (although with hindsight we include some 
next higher order terms which will be needed later).  Using~\eqn{3.870.1} we 
find
\bal{3.870.6}
b_1 &=& ck/z + kz + O(z^3) \nn
b_{\{{}^2_3\}} &=& \mp k|z|^{1\mp c}|z_i|^{\pm c} \nn
b_4 &=& (c+1)k/z_i + kz^3/3 + O(z^5)
\ea
and for the $a_{ij}$'s we need the following expressions, calculated 
from~\eqn{3.870.2.1-3}:
\bal{3.843-1}
{\im[X(z)X^\ast(\zeta)]\over \im X(z)} &\simeq& {(1-z/\zeta)\cos(\zeta-z)-
(z+1/\zeta)\sin(\zeta-z)\over \cos z + z\sin z} \nn
{\im[X(\zeta)]\over \im X(z)} &\simeq& {{\cos \zeta\over \zeta}+\sin\zeta
\over {\cos z\over z}+\sin z}
\ea
\bel{3.866.1-4.5}
\exp\left(\hat \gamma_0\int_{z_i}^\zeta {c^2(\zeta'')\over M}d\zeta''\right) = 
(\zeta/z_i)^{-c}\;\;\; ;\;\;\;
\exp\left(-\hat \gamma_0\int_{\zeta}^z {c^2(\zeta'')\over M}d\zeta''\right) = 
(z/\zeta)^c
\ee

\subsection {Entropy}

\subsubsection*{High temperature}

We begin by writing
\bal{3.856-0}
\nu &=& 4cc^2(s)T\delta(s-s') \nn
&=& {-4c k^2 T\over\zeta}\;\delta(\zeta-\zeta')
\ea
We calculate $a_{11}$ here and leave the details of $a_{12}, a_{22}$ to 
appendix~\ref{cal-of-aijs-in-chapter-entropy}.
First, \eqn{3.870.1} gives
\bal{3.866.1-5}
a_{11} &=& {1\over 2k^2} \int_{z_i}^z d\zeta\int_{z_i}^z d\zeta'\;
\left(\zeta\over z_i\right)^{-c}{\im [X(z)X^\ast(\zeta)]\over \im X(z)}
\;{4c k^2T\over -\zeta}\;\delta(\zeta-\zeta')
\left(\zeta'\over z_i\right)^{-c}{\im [X(z)X^\ast(\zeta')]\over\im X(z)}
\nn
&=& 2cT \int_{z_i}^z d\zeta\;\left(\zeta\over z_i\right)^{-2c}
\left({\im [X(z)X^\ast(\zeta)]\over \im X(z)}\right)^2\;{1\over -\zeta}
\ea
We wish to investigate the dependence of the $a_{ij}$'s on $z$ as $z\to 0$,
and so we now separate each integral into a sum of two parts.  The first is
gotten by integrating in to some constant $\lambda$ close to $z$, while the 
second integral contains the $z$ upper limit:
\bel{3.843.1.8}
a_{11} = 2cT \left[\inta + \intb \right]d\zeta \;\left(\zeta\over z_i
\right)^{-2c}\left({\im [X(z)X^\ast(\zeta)]\over \im X(z)}\right)^2\;
{1\over -\zeta}
\ee
It's only necessary to work to leading order in $z$.  We need the following 
expressions: when only $z\approx 0$ we have the $z$ dependence in the 
integrands as
\bal{3.843.8-1}
{\im[X(z)X^\ast(\zeta)]\over \im X(z)} &=& \cos\zeta-\sin\zeta/\zeta + O(z^2)
\equiv f_1(\zeta)+ O(z^2) \nn
{\im[X(\zeta)]\over \im X(z)} &\simeq& z(\cos\zeta/\zeta+\sin\zeta)\equiv 
zf_2(\zeta)
\ea
while if both $z, \zeta \approx 0$ then to leading order
\bel{3.843.8-2}
{\im[X(z)X^\ast(\zeta)]\over \im X(z)} \simeq (-\zeta^2+z^3/\zeta)/3 
\;\;\;,\;\;\;
{\im[X(\zeta)]\over \im X(z)} \simeq z/\zeta
\ee
We are now in a position to write
\bal{3.866.1-5-again}
a_{11} &\propto& c T \left[\int_{z_i}^\lambda d\zeta\;|\zeta|^{-2
c-1}f_1^2(\zeta)+\int_\lambda^z d\zeta\;|\zeta|^{-2c-1}
(-\zeta^2+z^3/\zeta)^2/9\right] \nn
&=& c T \left(O(1) + O|z|^{-2c+5}\right) \nn
&=& c T \;O(1)
\ea
since we have taken $c$ to be small.
A similar approach gives the following results for $a_{12}, a_{22}$
(details can be found in appendix~\ref{cal-of-aijs-in-chapter-entropy}):
\bel{3.866.2-and-3}
a_{12} = c T \;O|z|^{c+1} \;\;\;,\;\;\;
a_{22} = c T \;O(1)
\ee
Since $T$ is large, $a_{11}$ dominates $D$ while $a_{22}$ dominates $A$; 
so we have
\bel{3.866.4}
A \to a_{22} \;\;\;,\;\;\;
B \to -b_1/2 \;\;\;,\;\;\;
C \to {b_2^2\over 16 a_{11}}
\ee
These of course have the same form as for the static oscillator case, although 
it's by no means clear whether such a fact could have been deduced from the 
general expressions for the $a_{ij}$'s.  We now have
\be
S_{\it lin} \to {-|b_2|\over 4\sqrt{a_{11}a_{22}}} = O|z|^{1-c}
\ee
and using (\ref{ent-3-3}, \ref{3.870.8-5}) we can write
\bel{3.870.7-5}
S \to (1-c)r + {\rm constant}
\ee

\subsubsection*{Finite temperature}

Here we leave the frequency integration until last as was done for the static 
oscillator.  The integrals can then be done in the same way as in the last 
section, although some subtleties are present in this case (see 
appendix~\ref{cal-of-aijs-in-chapter-entropy}).  We finally obtain
\be
a_{11} = ck\;O(1) \;\;\;,\;\;\;
a_{12} = ck\;O|z|^{1/2} \;\;\;,\;\;\;
a_{22} = ck\;O(z)
\ee
Again since we integrate over $\hat\omega$, $a_{11}$ will be large and so
dominate $D$, leading to
\bel{3.866.8}
A \to a_{22}-{a_{12}^2\over 4a_{11}} \;\;\;,\;\;\;
B \to -b_1/2 \;\;\;,\;\;\;
C \to {b_2^2\over 16 a_{11}}
\ee
and so
\be
S_{\it lin} \to O|z|^{1/2-c}
\ee
Then with (\ref{ent-3-3}, \ref{3.870.8-5}) we have
\be
S \to (1/2-c)r + {\rm constant}
\ee

\section{Discussion}

In the last two sections we calculated the entropy of two
physical and exactly solvable squeezed systems; an inverted
harmonic oscillator and a scalar field mode evolving in a de Sitter 
inflationary universe. Our aim was to compare these results, based on our 
rigorous quantum open system framework, with that of the previous more
ad hoc approaches described in the introduction.
We must bear in mind that these previous results referred to a field 
mode that could be split into 2 independent sine and cosine (standing wave) 
components. We should therefore expect a
result of $S=r$ (rather than $2r$) if we are to agree with previous work.

For the inverted oscillator, in both temperature regimes with low
coupling, we obtained $S\to r +\mbox{\rm constant}$.
In the de Sitter case, the high temperature result is $S\to (1-c)r +
\mbox{\rm constant}$.  These three examples certainly do confirm the ad
hoc approaches to calculating entropy that have been used by others.
However at lower temperatures the de Sitter entropy is $S\to (1/2-c)r +
\mbox{\rm constant}$.  This last result requires us to look more closely
at $A$ and $C$ which together give the entropy. 

From~\eqnn{3-2.5}{ent-3-3}, 
and neglecting the added constants which are always implied,
we find that in the high squeezing limit the entropy behaves as
$$
S \to {1\over 2} \ln A - {1\over 2} \ln C.
$$
When the system-environment coupling is small, all of the above cases give
$-1/2 \ln C \to r$, which is the expected result. 
The dominant contribution to $C$ always comes from $b_2$ in the high 
squeezing limit. This parameter is determined by the squeezing of the system 
and is essentially independent on the nature of the environment and its 
coupling to the system. We can therefore conclude that the $\ln C$
contribution to the entropy represents entropy intrinsic to the squeezed
system itself. This is in agreement with the previous results and should
also be true quite generally for squeezed systems. However these results
cannot but fail to take into account the contributions to the entropy from the
$\ln A$ term. This contribution 
is determined by the $a_{ij}$ factors which strongly depend on the nature 
of the environment and its coupling to the system. 
There is, {\it a priori}, no reason to expect this contribution to be small,
a point illustrated by our finite temperature de Sitter example
for which we found $1/2 \ln A \to -r/2$. This highlights the 
danger in using the previous ad hoc approaches to entropy of squeezed systems.
The critical point is that the entropy of a system depends not only 
on the system itself but also on the nature of the environment it is
coupled to.

In conclusion, approaching the problem of entropy and uncertainty
from the open system viewpoint as we have demonstrated improves on the
earlier work in that it makes explicit how their dependence on the
coarse-graining of the environment and the system-environment couplings.
It also clarifies the relation between quantum and classical
descriptions -- it is through decoherence that the quantum field becomes
classical~\cite{CH94,AngZur}.
These issues are important as they rest at the
foundation of statistical and quantum mechanics.
(For a discussion of the deeper meaning of the dependence of persistent
structures on coarse-graining, see~\cite{HuSpain}.)\\

\noindent {\bf Acknowledgement}
DK thanks the Australian Vice Chancellors' Committee
for its financial support.
BLH acknowledges support from the U. S. National Science Foundation
under grants PHY94-21849, and the General Research Board of the
Graduate School of the University of Maryland.  BLH and AM enjoyed
the hospitality of the physics department of the Hong Kong University of
Science and Technology in the spring of 1995 when part of this work was done.

\newpage
\appendix
 
\section{Details for section \protect \ref{superpropagator}}

\subsection{Calculating $u_1 \rightarrow v_2$}

Now we are in a position to solve~(\ref{qbm3-3.5}, \ref{qbm3-3.6}) for
$u_1 \rightarrow v_2$.  First consider~\eqn{qbm3-3.5}.
We treat the integral of a delta function and its derivative in the following
way: use a smooth step function (i.e.\ $\theta(0) \equiv 1/2$) to write
($x_1>x_0$)
\bal{3.601.5}
\int_{x_0}^{x_1}f(x)\delta(x-a)\;dx &\equiv& \;\;\;\,f(a) \;\theta(x_1-a)\;
\theta(a-x_0) \\
\int_{x_0}^{x_1}f(x)\delta'(x-a)\;dx &\equiv& -f'(a)\;\theta(x_1-a)\;
\theta(a-x_0)
\ea
These relations can easily be proved by checking the five cases individually,
of $a < x_0$, $a = x_0$, $x_0<a<x_1$ etc.
Note that treating the delta function in this `smoothed' way eliminates the
need for the frequency renormalisation in~\cite{PHZ}.
This smoothing essentially just defines $\int_0^\infty \delta(x) dx =1/2$
(see e.g.~\cite{von-neumann} for a discussion of this).

Hence \eqn{qbm3-3.5} together with \eqn{mu-and-nu} becomes (with $u$ being 
either $u_1$ or $u_2$)
\bel{3.601.9-3}
\ddot u(s)+\left({\dot M\over M} + {2\gamma_0c^2\over M}\right)\dot u+
\left(\Omega^2+{\dot M{\cal E}\over M} + \dot{\cal E}+{2\gamma_0c\dot c\over 
M}\right) u=0
\ee
Now define $\tilde u$ by
\bel{3.601.9-4}
\tilde u \equiv u \exp\left[\gamma_0\int_{t_i}^s {c^2(s')\over M(s')}ds'\right]
\ee
in which case it follows that
\bel{3.601.10-1}
\ddot {\tilde u} +{\dot M\over M}\dot {\tilde u} + \left(\Omega^2+{\dot M{\cal 
E}\over M} + \dot{\cal E}
-{\gamma_0^2c^4\over M^2}\right)\tilde u =0
\ee

Comparing with \eqn{5-1}, we recognise this as just
the equation of motion of an oscillator with
mass $M$, cross term ${\cal E}$ and an effective frequency
\bel{3.601.10-2}
\Omega_{\it eff}^2 \equiv \Omega^2 -{\gamma_0^2c^4\over M^2}
\ee

So, we are in a position to describe our system in terms of an equivalent
system.  Hence we know a solution for $\tilde u(s)$---it is the sum $X$ of the 
Bogoliubov coefficients for this new system.  So we write 
(with $g_1, g_2$ constants to be determined)
\be
u(s) = \exp\left[-\gamma_0\int_{t_i}^s {c^2\over M}ds'\right]
[g_1 X(s) + g_2 X^{\ast}(s)]
\ee
By including the boundary conditions for $u_1$ and $u_2$ we obtain
\bal{3.850}
u_1(s) &=& \exp\left[-\gamma_0\int_{t_i}^s {c^2\over M}ds'\right]
\;{\im [X(t)X^{\ast}(s)]\over \im X(t)}\nn
&&\nn
u_2(s) &=& \exp\left[\gamma_0\int_s^t {c^2\over M}ds'\right]
\;{\im X(s)\over \im X(t)}
\ea
This tying in of the propagator formalism to the language of squeezed states
(such as Bogoliubov coefficients) will be very useful for relating
the entropy of a field mode to its squeeze parameter~$r$.

In the same way that we solved \eqn{qbm3-3.5}, eqn~\eqn{qbm3-3.6} becomes
\bel{3.850.1-5}
\ddot v(s)+\left({\dot M\over M} - {2\gamma_0c^2\over M}\right)\dot v +
\left(\Omega^2+{\dot M{\cal E}\over M} + \dot{\cal E}-{2\gamma_0c\dot c\over 
M}\right) v=0
\ee
Now write
\bel{3.850.2-0}
\tilde v \equiv v \exp\left[-\gamma_0\int_{t_i}^s {c^2\over M}ds'\right]
\ee
and just as for the case of $u$ we have
\bel{3.850.2-1}
\ddot {\tilde v} +{\dot M\over M}\dot {\tilde v} + \left(\Omega^2+{\dot M{\cal 
E}\over M} + \dot{\cal E}-
\frac{\gamma_0^2c^4}{M^2}\right)
\tilde v =0
\ee
So now $v_1$ and $v_2$ can also be written as combinations of $X$ and $X^\ast$.
Including the boundary conditions we eventually obtain
\bal{3.850.3}
v_1(s) &=& \exp\left[\gamma_0\int_{t_i}^s {c^2\over M}ds'\right]
\;{\im [X(t)X^{\ast}(s)]\over \im X(t)}\nn
&&\nn
v_2(s) &=& \exp\left[-\gamma_0\int_s^t {c^2\over M}ds'\right]
\;{\im X(s)\over \im X(t)}
\ea

\subsection{Calculating $a_{11} \rightarrow b_4$}

To facilitate our calculations we introduce dimensionless parameters for time
\bal{3.614.6-1}
z &\equiv& \kappa t \;\;\;,\;\;\; \zeta \equiv \kappa s \nn
X(z) &\equiv& X(t) \;\;{\rm etc.}
\ea
and a carat will denote division by $\kappa$, e.g.\ $\hat\gamma_0 = 
\gamma_0/\kappa$.  Note that $t$ is the lagrangian time, which isn't 
necessarily cosmic.

Now we are able to calculate the propagator.  
Making use of (\ref{qbm3-3.12}, \ref{qbm3-3.11}) we obtain
\begin{eqnarray*}
a_{11}(z,z_i) &=&{1\over 2\kappa^2}\int_{z_i}^z d\zeta\int_{z_i}^z
d\zeta'\;
e^{\hat\gamma_0\int_{z_i}^\zeta {c^2\over M}d\zeta''}
{\im[X(z)X^\ast(\zeta)]\over\im X(z)}\;\nu(\zeta,\zeta')\;
e^{\hat\gamma_0\int_{z_i}^{\zeta'} {c^2\over M}d\zeta''}
{\im[X(z)X^\ast(\zeta')]\over\im X(z)} \nn
&& \nn
a_{12} &=& {1\over \kappa^2}\int_{z_i}^z d\zeta\int_{z_i}^z
d\zeta'\;
e^{\hat\gamma_0\int_{z_i}^\zeta {c^2\over M}d\zeta''}
{\im[X(z)X^\ast(\zeta)]\over\im X(z)}\;\nu(\zeta,\zeta')\;
e^{-\hat\gamma_0\int_{\zeta'}^z {c^2\over M}d\zeta''}
{\im X(\zeta')\over\im X(z)} \nn
&& \nn
a_{22} &=& {1\over 2\kappa^2}\int_{z_i}^z d\zeta\int_{z_i}^z
d\zeta'\;
e^{-\hat\gamma_0\int_\zeta^z {c^2\over M}d\zeta''}
{\im X(\zeta)\over\im X(z)}\;\nu(\zeta,\zeta')\;
e^{-\hat\gamma_0\int_{\zeta'}^z {c^2\over M}d\zeta''}
{\im X(\zeta')\over\im X(z)}
\end{eqnarray*}
\bal{3.870.1}
b_1(z,z_i) &=& -\hat\gamma_0\,\kappa c^2(z)  +\kappa M(z){\im X'(z)\over 
\im X(z)} +M(z){\cal E}(z)\nn
&&\nn
b_{\{^2_3\}} &=& {\mp \kappa \,e^{\pm\hat\gamma_0
\int_{z_i}^z{c^2\over M}d\zeta}\over\im X(z)} \nn
&&\nn
b_4 &=& -\hat\gamma_0\,\kappa c^2(z_i) + \kappa \,{\re X(z)\over \im X(z)}+
M(z_i){\cal E}(z_i)
\ea

\section {Entropy of a static oscillator in a thermal bath}
The lagrangian for the static oscillator with unit mass is given by
\bel{3.611-1}
L = {1\over 2}\left(\dot x^2-k^2x^2\right)
\ee
{}From~\eqn{3.601.10-2} with $M=c=1$ the effective frequency is
\bel{3.611-2}
\Omega_{\it eff}^2 = k^2-\gamma_0^2 \equiv \kappa^2
\ee
Then the equation of motion for $X$ is, from~\eqn{5-1} with $\Omega\to
\Omega_{\it eff}$
\bel{3.611-3-and-4}
\ddot X+\kappa^2 X =0
\ee
\be
X(0) = 1\;\;\;,\;\;\;
\dot X(0) = -i\kappa
\ee
which leads to
\bel{3.611-5}
X(z) = e^{-iz}
\ee
with $z = \kappa t$.  Then
\bel{3.611-6-and-7}
{\im [X(z)X^\ast(\zeta)]\over \im X(z)} = {\sin(z-\zeta)\over \sin z}
\;\;\;,\;\;\;
{\im X(\zeta)]\over \im X(z)} = {\sin\zeta\over \sin z}
\ee
\bel{3.611-8}
e^{\hat\gamma_0\int_{z_i}^\zeta {c^2\over M}d\zeta''} = e^{\hat \gamma_0\zeta}
\;\;\;,\;\;\;
e^{-\hat\gamma_0\int_\zeta^z {c^2\over M}d\zeta''}=e^{-\hat \gamma_0(z-\zeta)}
\ee
with noise for $T\rightarrow\infty$ being white:
\bel{3.611-9}
\nu(\zeta,\zeta') = 4\kappa\gamma_0 T\delta(\zeta-\zeta') 
\ee
Then $a_{11}\to b_4$ follow:
\ba
a_{11} &=& {T\over \sin^2z}\cdot{e^{2\hat\gamma_0z}-1-\hat\gamma_0\sin2z-
\hat\gamma_0^2(1-\cos2z)\over 2(1+\hat\gamma_0^2)} \nn
&&\nn
a_{12} &=& {2T\over \sin^2z}\cdot{-\cos z\sh \hat\gamma_0z + 
\hat\gamma_0\sin z\ch \hat\gamma_0z\over 1+\hat\gamma_0^2} \nn
&&\nn
a_{22} &=& {T\over \sin^2z}\cdot{-e^{-2\hat\gamma_0z}+1-\hat\gamma_0\sin2z+
\hat\gamma_0^2(1-\cos2z)\over 2(1+\hat\gamma_0^2)} \nonumber
\ea
\bel{3.611.1-and-3.611.2}
b_{\{^1_4\}} = \kappa (-\hat\gamma_0\pm\cot z) \;\;\;,\;\;\;
b_{\{^2_3\}} = {\pm\kappa \;e^{\pm \hat\gamma_0z}\over \sin z}
\ee

To evaluate $S$, we need $A$ and
$C$; in turn for these we need $a_{11} \rightarrow b_4$.  These are 
calculated from~\eqn{3.870.1}.\footnote{
Various notations exist describing these results; see for example
\cite{Mat,ZHP,HuZhaUncer}.
To compare with~\cite[eqn~2.2.7]{HuZhaUncer} is a matter of carefully 
transcribing the notation; key things to note are that $X \equiv \Sigma$, 
$Y\equiv -\Delta$; here we have taken $x_0 =p_0 =0$; 
\cite[eqn~2.2.6c]{HuZhaUncer} should have an $a_{11}$ in place of the $a_{22}$; 
the $b_i$'s in~\cite{HM2} are written explicitly in~\cite{HuZhaUncer} 
via~\cite[3.11]{HM2};~\cite[$a_{12}$]{HM2}
equals~\cite[$a_{12}+a_{21}$]{HuZhaUncer};~\cite[$\gamma_0$]{HM2} equals 
\cite[$\gamma_0/2$]{HuZhaUncer}.
}

The oscillator is assumed to be initially in its ground state
\bel{3.610.3-1}
\psi(x,0) \propto \exp{-x^2\over 4\sigma^2}
\ee
so that its density matrix is
\bel{3.610.3-2}
\rho(x\,x'\,0) \propto \exp{-x^2-x'^2\over 4\sigma^2}
\ee
and in \eqn{3.723-1} we have
\bel{3.613.2-1}
\xi = {1\over 4\sigma^2}\;\;\;,\;\;\; \chi=0
\ee
The reduced density matrix evolves into \eqn{3-0.5}, with
\bal{3.613.2-2}
A &=& a_{22}+{1\over D}\left\{\left[{1\over 8\sigma^2}+a_{11}\right]b_3^2
+a_{12}b_3b_4-{a_{12}^2\over 2\sigma^2}\right\} \nn
B &=& {-b_1\over 2} +{b_2\over 2D} \left\{b_3b_4 -{a_{12}\over\sigma^2}\right\}
\nn
C &=& {b_2^2\over 8D\sigma^2} \nn
D &=& {1\over 4\sigma^4} + {2a_{11}\over \sigma^2}+b_4^2
\ea
It's by no means trivial to show that the entropy calculated using these 
expressions does indeed tend toward~\eqn{2.05-5}, and in particular the
$\csc z$ terms in the $a_{ij}$'s and $b_i$'s mean their values can diverge
depending
on the time.  But this divergence cancels out when physical quantities are
measured, as we can see by verifying numerically that our entropy really does
tend toward the usual asymptotic value at late times (Fig.~1).

\section{Calculation of $a_{ij}$'s in section \protect\ref{de-sitter-section}}
\Label{cal-of-aijs-in-chapter-entropy}

\subsection*{de Sitter with high temperature}

Here we evaluate the $a_{ij}$'s leading 
to~\eqn{3.866.2-and-3}.  We are using the following small $z, \zeta$ 
approximations:
\bal{3.843.8}
{\im [X(z)X^\ast(\zeta)]\over \im X(z)} &\stackrel{z\to 0}{\longrightarrow}&
\cos \zeta -\sin\zeta/\zeta + O(z^2) \equiv f_1(\zeta) + O(z^2) \nn
&\stackrel{z,\zeta\to 0}{\longrightarrow}& (-\zeta^2+z^3/\zeta)/3 \nn
{\im X(\zeta)\over \im X(z)} &\stackrel{z\to 0}{\longrightarrow}&
z(\cos \zeta/\zeta +\sin\zeta)\equiv zf_2(\zeta) \nn
&\stackrel{z,\zeta\to 0}{\longrightarrow}& z/\zeta
\ea
Firstly, 
\bal{3.866.1-6}
a_{12} &=& {1\over k^2} \int_{z_i}^z d\zeta\int_{z_i}^z d\zeta'\;
\left(\zeta\over z_i\right)^{-c}{\im [X(z)X^\ast(\zeta)]\over \im X(z)}
\;{4c k^2T\over -\zeta}\;\delta(\zeta-\zeta')
\left(z\over\zeta'\right)^c{\im X(\zeta')\over\im X(z)} \nn
&=& 4cT \int_{z_i}^z d\zeta\;\left(\zeta\over z_i\right)^{-c}
{\im [X(z)X^\ast(\zeta)]\over \im X(z)}\;{1\over -\zeta}\;
\left(z\over\zeta\right)^c{\im X(\zeta)\over\im X(z)} \nn
&\propto& c T |z|^c\left[\int_{z_i}^\lambda d\zeta\;
|\zeta|^{-2c-1}f_1(\zeta)\,z\,f_2(\zeta)+\int_\lambda^z d\zeta\;
|\zeta|^{-2c-1}(-\zeta^2+z^3/\zeta)\;{z\over 3\zeta}\right] \nn
&=& c T \;O|z|^{c+1}
\ea
provided $c < 1/2$.  Finally,
\bal{3.866.1-7}
a_{22} &=& {1\over 2k^2} \int_{z_i}^z d\zeta\int_{z_i}^z d\zeta'\;
\left(z\over\zeta\right)^c{\im X(\zeta)\over \im X(z)}
\;{4c k^2T\over -\zeta}\;\delta(\zeta-\zeta')
\left(z\over\zeta'\right)^c{\im X(\zeta')\over\im X(z)} \nn
&=& 2cT \int_{z_i}^z d\zeta\;\left(z\over\zeta\right)^{2c}\left(
{\im X(\zeta)\over \im X(z)}\right)^2\;{1\over -\zeta} \nn
&\propto& c T |z|^{2c}\left[\int_{z_i}^\lambda d\zeta\;
|\zeta|^{-2c-1}f_2^2(\zeta)\,z^2+\int_\lambda^z d\zeta\;
|\zeta|^{-2c-1}z^2/\zeta^2\right] \nn
&=& c T \;O(1)
\ea

\subsection*{de Sitter with finite temperature}

We leave the frequency integration until last:
\bal{3.864-1}
\nu &=& {2c\over\pi}{1\over\sqrt{ss'}}\int_0^\infty\omega \coth {\omega
\over 2T}\cos\omega(s-s')\;d\omega \nn
&=& {2c\over\pi}{k^3\over \sqrt{\zeta\zeta'}}\int_0^\infty 
d\hat\omega\; \hat\omega \coth {\hat\omega\over 2T}\cos\hat\omega(\zeta-\zeta')
\ea
The $a_{ij}$'s are
\bal{3.864}
a_{11} &=& z_i^{2c}{ck\over\pi}\int_0^\infty d\hat\omega\;
\hat\omega\coth {\hat\omega k\over 2T}\times \nn
&&\underbrace{
\int_{z_i}^z d\zeta\int_{z_i}^z d\zeta'\;
|\zeta|^{-c-1/2}{\im [X(z)X^\ast(\zeta)]\over \im X(z)}
\;\cos\hat\omega(\zeta-\zeta')\;
|\zeta'|^{-c-1/2}{\im [X(z)X^\ast(\zeta')]\over \im X(z)}}_{\equiv 
I_{11}} \nn
a_{12} &=& (z_iz)^c{2ck\over\pi}\int_0^\infty d\hat\omega\;
\hat\omega\coth {\hat\omega k\over 2T}\times \nn
&&\underbrace{
\int_{z_i}^z d\zeta\int_{z_i}^z d\zeta'\;
|\zeta|^{-c-1/2}{\im [X(z)X^\ast(\zeta)]\over \im X(z)}
\;\cos\hat\omega(\zeta-\zeta')\;
|\zeta'|^{-c-1/2}{\im X(\zeta')\over \im X(z)}}_{\equiv I_{12}} \nn
a_{22} &=& z^{2c}{ck\over\pi}\int_0^\infty d\hat\omega\;
\hat\omega\coth {\hat\omega k\over 2T}\times \nn
&&\underbrace{
\int_{z_i}^z d\zeta\int_{z_i}^z d\zeta'\;
|\zeta|^{-c-1/2}{\im X(\zeta)\over \im X(z)}
\;\cos\hat\omega(\zeta-\zeta')\;
|\zeta'|^{-c-1/2}{\im X(\zeta')\over \im X(z)}}_{\equiv I_{22}}
\ea
Using the expressions from (\ref{3.843.8}, \ref{3.866.1-4.5}) the first of 
the inner integrals becomes
\bal{3.866.5-2}
I_{11} &=& \inta d\zeta\;|\zeta|^{-c-1/2}f_1(\zeta)\left[
\inta d\zeta'\;\cos\hat\omega(\zeta-\zeta')\;|\zeta'|^{-c-1/2} 
f_1(\zeta')\right. \nn
&&{}+\left.
\intb d\zeta'\;\cos\hat\omega\zeta\;|\zeta'|^{-c-1/2}(-\zeta'^2+z^3/
\zeta')/3\right] \nn
&&{}+\intb d\zeta\;|\zeta|^{-c-1/2}(-\zeta^2+z^3/\zeta)/3\left[
\inta d\zeta'\;\cos\hat\omega\zeta' \;|\zeta'|^{-c-1/2}f_1(\zeta')
\right. \nn
&&{}+\left.
\intb d\zeta'\;\cos\hat\omega(\zeta-\zeta')\;|\zeta'|^{-c-1/2}
(-\zeta'^2+z^3/\zeta'/3)\right]
\ea
We now have a difficulty.  In order to get a reasonably useful analytic
result, it will be an advantage to replace the $\cos\hat\omega(\zeta-\zeta')$
term in the fourth integral above by something simpler.
We will have competition between $\hat\omega$ increasing in the frequency
integral versus $z$ decreasing in time.
Suppose then we use a frequency cutoff $\omega_{\it max}$.  In that
case we can approximate $\cos\hat\omega(\zeta-\zeta')$ for $\zeta,\zeta'
\approx 0$ by choosing $\hat\omega_{\it max}$ such that $\cos\hat
\omega(\zeta-\zeta') \approx 1$ in the fourth integral.  This will be true
provided
\be
\hat\omega_{\it max} \ll -1/\lambda
\ee
However now we don't expect our result to necessarily agree with the high $T$ 
result found
in~\eqn{3.866.1-5-again}, since there we had taken $\hat\omega_{\it max}
\to \infty$, which was made possible by the use of the delta function.

At this point we refer to the discussion of the high temperature limit in 
\cite{KoksPhD}. There it is shown that the high temperature
(delta function) regime is that for which $\omega_{\it max} \ll T$
\underline{and} $\omega_{\it max} \to\infty$.  This absence of a cutoff in the
high temperature limit is usually not stressed, but it forms the most
relevant fact here.  In general we must impose a cutoff for all finite $T$
values, otherwise the frequency integral is not well defined---unless 
$T\to\infty$.  So we conclude that the regime for which our analysis is valid
here is $T \ls \omega_{\it max}$.

With the last cosine set equal to 1 as before, these integrals are all $O(1)$
and therefore so is $a_{11}$.  Next:
\bal{3.866.6-6}
I_{12} &=& \inta d\zeta\;|\zeta|^{-c-1/2}f_1(\zeta)\left[
\inta d\zeta'\;\cos\hat\omega(\zeta-\zeta')\;|\zeta'|^{-c-1/2} 
f_2(\zeta')\,z\right. \nn
&&{}+\left.
\intb d\zeta'\;\cos\hat\omega\zeta\;|\zeta'|^{-c-1/2}z/\zeta'\right] \nn
&&{}+\intb d\zeta\;|\zeta|^{-c-1/2}(-\zeta^2+z^3/\zeta)/3\left[
\inta d\zeta'\;\cos\hat\omega\zeta' \;|\zeta'|^{-c-1/2}f_2(\zeta')\,z
\right. \nn
&&{}+\left.\intb d\zeta'\;\cos\hat\omega(\zeta-\zeta')\;|\zeta'|^{-c-
1/2}z/\zeta'\right]
\ea
Evaluating these integrals gives $I_{12} = O|z|^{-c+1/2}$ so that
$a_{12} = O|z|^{1/2}$.  Lastly,
\bal{3.866.7-10}
I_{22} &=& \inta d\zeta\;|\zeta|^{-c-1/2}f_2(\zeta)\,z\left[
\inta d\zeta'\;\cos\hat\omega(\zeta-\zeta')\;|\zeta'|^{-c-1/2} 
f_2(\zeta')\,z\right. \nn
&&{}+\left.
\intb d\zeta'\;\cos\hat\omega\zeta\;|\zeta'|^{-c-1/2}z/\zeta'\right] \nn
&&{}+\intb d\zeta\;|\zeta|^{-c-1/2}z/\zeta\left[
\inta d\zeta'\;\cos\hat\omega\zeta' \;|\zeta'|^{-c-1/2}f_2(\zeta')\,z
\right. \nn
&&{}+\left.\intb d\zeta'\;\cos\hat\omega(\zeta-\zeta')\;|\zeta'|^{-c-
1/2}z/\zeta'\right] \nn
&=& O|z|^{-2c+1}
\ea
so that $a_{22}=O(z)$.


\end{document}